\documentclass[aps,prb,reprint,twocolumn,superscriptaddress]{revtex4-1}

\usepackage{amsmath}
\usepackage{amssymb}
\usepackage{graphicx}
\usepackage{epstopdf}
\usepackage{multirow}
\usepackage[usenames,dvipsnames]{xcolor}

\usepackage{tabularx}
\usepackage{array}
\newcommand{\deftab}[2]{\begin{tabularx}{#1}{*{#2}{|>{\centering\arraybackslash}X}|}}

\newcommand\loz{L1$_0$ }

\newcommand\lot{L1$_2$ }
\newcommand\lozz{L1$_0$}
\newcommand\lott{L1$_2$}

\newcommand\beq{\begin{equation}}
\newcommand\eeq{\end{equation}}
\newcommand\unbeq{\end{equation}}
\newcommand\bea{\begin{eqnarray}}
\newcommand\eea{\end{eqnarray}}

\newcommand\dH{\Delta H}
\newcommand\mdH{$\Delta H$ }

\begin{document}

\title{Phase stability, ordering tendencies, and magnetism in single-phase fcc Au-Fe nanoalloys}

\author{I. A. Zhuravlev}

\affiliation{Department of Physics and Astronomy and Nebraska Center for Materials and Nanoscience, University of Nebraska--Lincoln, Lincoln, Nebraska 68588, USA}

\author{S. V. Barabash}
\affiliation{Department of Physics and Astronomy and Nebraska Center for Materials and Nanoscience, University of Nebraska--Lincoln, Lincoln, Nebraska 68588, USA}
\affiliation{Intermolecular Inc., San Jose, California 95134, USA}

\author{J. M. An}

\affiliation{Department of Physics and Astronomy and Nebraska Center for Materials and Nanoscience, University of Nebraska--Lincoln, Lincoln, Nebraska 68588, USA}

\author{K. D. Belashchenko}

\affiliation{Department of Physics and Astronomy and Nebraska Center for Materials and Nanoscience, University of Nebraska--Lincoln, Lincoln, Nebraska 68588, USA}

\date{\today}

\begin{abstract}
Bulk Au-Fe alloys separate into Au-based fcc and Fe-based bcc phases, but L1$_0$ and L1$_2$ orderings were reported in single-phase Au-Fe nanoparticles. Motivated by these observations, we study the structural and ordering energetics in this alloy by combining density functional theory (DFT) calculations with effective Hamiltonian techniques: a cluster expansion with structural filters, and the configuration-dependent lattice deformation model. The phase separation tendency in Au-Fe persists even if the fcc-bcc decomposition is suppressed. The relative stability of disordered bcc and fcc phases observed in nanoparticles is reproduced, but the fully ordered L1$_0$ AuFe, L1$_2$ Au$_3$Fe, and L1$_2$ AuFe$_3$ structures are unstable in DFT. However, a tendency to form concentration waves at the corresponding [001] ordering vector is revealed in nearly-random alloys in a certain range of concentrations. This incipient ordering requires enrichment by Fe relative to the equiatomic composition, which may occur in the core of a nanoparticle due to the segregation of Au to the surface. Effects of magnetism on the chemical ordering are also discussed.
\end{abstract}

\maketitle

\section{Introduction}

The theory of phase stability in crystalline alloys usually relies on the assumption that the set of alloy configurations is in one-to-one correspondence with the set of decorations of the underlying parent lattice, which allows one to reduce the problem to an Ising model on that lattice. \cite{deFontaine,SanchezCE} While this assumption is justified in many substitutional alloy systems, it becomes problematic if two or more different parent lattices (for example, bcc and fcc) compete with each other. In this case some configurations of an Ising model for the given lattice may correspond to dynamically unstable (i.e., non-existent) physical configurations, complicating both the construction of an appropriate effective Hamiltonian and the prediction of thermodynamic properties. This situation is most likely to occur in systems undergoing phase separation into two phases with different crystal lattices. Due to the failure of the standard alloy-theoretical methods, the phase stability in such systems remains largely unexplored.

The equilibrium bulk phase diagram \cite{ASM} shows that Au-Fe alloys phase-separate and have no equilibrium ordered phases. However, this phase separation was not observed in nanoparticles. After low-temperature deposition, Au-Fe nanoparticles with 65\% Fe or more were found to have body-centered cubic (bcc) structure, while those with 53\% Fe or less were face-centered cubic (fcc).\cite{Shield,Shield2013,Shield2014} After the subsequent heat treatment, which included a recrystallizing high-temperature annealing followed by slow cooling, all nanoparticles with 33-79\% Fe were fcc.\cite{Shield,Shield2014}

Further, evidence of ordering was found in heat-treated fcc nanoparticles. \cite{Sato,Shield,Shield2014} In particular, nearly stoichiometric 5-nm AuFe nanoparticles had a tetragonally distorted fcc structure, and L1$_0$ superstructure peaks were identified in the Fourier-transformed high-resolution transmission electron microscopy (HRTEM) images. \cite{Shield} Near AuFe$_3$ and Au$_3$Fe compositions, \lot phases were found.\cite{Shield2014} We also note that L1$_0$-type AuFe phase was artificially fabricated by monolayer deposition, which suggests that it is metastable in a thin film geometry.\cite{Takanashi}

Phase separation in nanoparticles may be blocked or suppressed either thermodynamically or kinetically by several mechanisms. (1) The free energy gain from phase separation scales with the volume, and the cost of forming an interphase boundary with the cross-section of a nanoparticle. Therefore, phase separation may be suppressed below a certain size. (2) Large surface energyq of one phase may stabilize the other phase in small particles.\cite{Gruner} (3) Spinodal decomposition is kinetically suppressed as the particle size becomes comparable to the Cahn-Hilliard wavelength, which determines the fastest-growing concentration fluctuation in the bulk material.\cite{Hoyt} Mukherjee {\em{et al.}} have argued \cite{Shield2013-barrier} that thermodynamic suppression of phase separation may indeed be responsible for some of the experimental observations in Au-Fe nanoparticles.

\loz and \lot phases are commonly found in compound-forming alloys of Au (e.g., Cu-Au) and Fe (e.g., Fe-Pt), but compound formation is not expected in Au-Fe, since the initial electron density mismatch between Fe and Au
is too large to be overcome by the relatively small charge transfer, according to the conventional metallurgical models.\cite{Miedema} This view of an inherent phase separation tendency in Au-Fe has been challenged by the theoretical \cite{Ling} and experimental \cite{Anderson-SRO,Fratzl-SRO} reports of coexisting phase-separation and ordering tendencies, which manifest themselves through short-range order, in disordered Au-rich Au-Fe alloys.
Only Au-rich Au-Fe alloys have been examined theoretically, and the contribution from strain-induced interaction was neglected.\cite{Ling} On the other hand, the structure of the ordered compounds is not necessarily inherited from the ordering tendencies in the random alloy, as exemplified by Ni-V and Pd-V alloys. \cite{Wolverton-SROLRO}

In this paper we analyze the phase stability of Au-Fe alloys, particularly as it relates to the experimental observations for AuFe nanoparticles. Since the size of the nanoparticles investigated in Ref.\ \onlinecite{Shield} is much larger than the metallic screening length, we assume no direct influence of the surface on the ordering tendencies in the particle core. The surface may, however, affect the ordering tendencies indirectly, by compressing the nanoparticle due to the surface tension, or by segregating one of the constituent elements to the surface and thereby depleting its core. Therefore, we first consider the structural hierarchy (including fcc-bcc stability and ordering tendencies) in bulk alloys, and then evaluate the influence of the indirect surface effects on these tendencies. We focus on the configurational ordering energetics, but also evaluate the possible role of different types of magnetic order. Our approach combines direct {\em{ab initio}} calculations with {\em{ab initio}}-based effective Hamiltonian techniques used to navigate the configurational space.

We find, with respect to the fcc-bcc stability:
(a) bulk energetics dictates that Au$_{1-x}$Fe$_x$ deposited at low temperatures should form the bcc phase at $x\gtrsim0.68$, consistent with experiment;
(b) under reasonable assumptions, annealing at $T\gtrsim 700$ K should transform the alloys with $x\lesssim 0.75$ to the fcc phase, consistent with experiment;
(c) the dynamic stability of both fcc and bcc lattices depends strongly on the atomic configuration, rather than just the concentration and temperature (as, for example, in Fe-Pd);
(d) although the energetics of fcc-bcc competition is drastically altered in some antiferromagnetic (AFM) structures (in particular, \loz becomes fcc-unstable in magnetic structures with antiparallel nearest neighbors), random deviations from perfect ferromagnetic (FM) order, up to and including the paramagnetic (PM) state, leave the \loz fcc-bcc transformation path qualitatively unchanged;
(e) immiscibility in Au-Fe alloys does {\em not} originate from the freedom to separate into fcc and bcc phases (as, e.g., in Fe-Ni), but, rather, both fcc and bcc alloys would already be immiscible in a wide concentration range.

We further find, with respect to the ordering tendencies:
(a) the assumption of {\em{full}} \loz ordering in AuFe and \lot in Au$_3$Fe
is in direct conflict with {\em{ab initio}} calculations, which indicate that {\em{lower-energy}}
fully ordered fcc-based structures exist at both compositions, including the so-called Z1 Au$_3$Fe and W2 AuFe, and, moreover, that some bcc-based AuFe structures are more stable than \loz AuFe at low temperatures;
(b) full L1$_2$ ordering is also unlikely for AuFe$_3$, where a number of dynamically unstable ordered structures are predicted by CE-SF to have lower energy than L1$_2$, and some bcc-based structures are also more stable than L1$_2$;
(c) in {\em{nearly-disordered}} AuFe$_3$, the ordering tendencies are characterized by the X-point ordering vector (consistent with \lot order), suggesting that the observed order type reflects partial ordering;
(d) nearly-disordered fcc alloys at AuFe stoichiometry do {\em not} exhibit ordering tendencies of \loz type and, moreover, are unstable with respect to spinodal decomposition;
(e) the lattice parameters of the experimentally observed \loz and \lott-ordered nanoparticles are much lower than the values predicted theoretically, suggesting a strong Fe enrichment of the nanoparticle core, with an additional contraction due to surface tension effects;
(f) Fe enrichment may induce \loz ordering tendencies in the nearly-disordered cores of nanoparticles with
a nominal AuFe composition, and may also spinodally stabilize them,
both effects being fully developed by the Au$_{1/3}$Fe$_{2/3}$ composition and beyond;
(g) additional contraction (due to surface tension or other effects) has a negligible effect on the ordering tendencies;
(h) magnetic disorder may qualitatively affect ordering; in particular,
(i) quenching of the nanoparticles annealed above the Curie temperature may reveal W-point ordering tendencies, such as ordering into
the CH structure; (j) Au$_3$Fe \lot is predicted to be FM, in contrast to earlier calculations \cite{Shield2014} suggesting antiferromagnetism.

We have not been able to reach conclusions about the ordering tendencies in the nearly-disordered Au-rich alloys, because our methodology predicts the random alloy to be dynamically unstable at those compositions.

The rest of the paper is organized as follows. Section\ \ref{sec:Methodology} reviews the key methodology, including the {\em{ab initio}} details in Sec.\ \ref{sec:DFT}, the cluster expansion with structural filters (CE-SF) in Sec.\ \ref{CEmethod}, and the configuration-dependent lattice deformation model (CLDM\cite{CLDM}) with its simplified version (S-CLDM) in Sec.\ \ref{CLDMmethod}. Only the key aspects of  the CE and CLDM are presented in the main text, while the technical details are given in the Appendices. Section\ \ref{sec:fccbcc} studies the fcc and bcc lattice stability in Au-Fe alloys, including miscibility and the general fcc/bcc competition (Sec.\ \ref{sec:fccbcc:random}), the dependence of the fcc/bcc transformations on configurational order (Sec.\ \ref{sec:fccbcc:ordered}) and on magnetic order (Sec.\ \ref{sec:magnetismBain}) along the Bain path. In Sec.\ \ref{sec:ordered}, we study the energetics of perfectly ordered structures, both in the fully relaxed geometry (Sec.\ \ref{sec:ordered:CE}) and subject to geometric relaxation constraints (as relevant to the CLDM construction,
Sec.\ \ref{sec:ordered:CLDM}). In Sec.\ \ref{sec:disordered}, we analyze how disordered alloys may develop ordering tendencies different from those found for the fully-ordered structures. The configurational energetics of nearly-random alloys is studied in Sec.\ \ref{sec:disordered:main}, the effects of surface segregation and surface tension on ordering in nanoparticles in Sec.\ \ref{sec:disordered:volume}, and the spinodal stability of fcc alloys in Sec.\ \ref{sec:disordered:spinodal}.
Sec.\ \ref{sec:magnetism} studies the effects of magnetic disorder on chemical ordering, and Sec.\ \ref{sec:Discussion} presents further discussion and conclusions. Finally, the Appendices summarize the technical details of CE and CLDM, and present a proof that striction has no effect on the ordering tendencies in a random alloy.

\section{Methodology}
\label{sec:Methodology}

To adequately model the phase stability in alloys, one needs to evaluate the energetic competition between many possible ordered structures. In Au-Fe alloys this task is complicated by the fcc-bcc competition, since some ordered fcc structures may by dynamically unstable and relax without a barrier towards a bcc structure (and vice versa). Moreover, such ``mixed-lattice'' alloys might lack any clustering tendency within the given (fcc or bcc) lattice type, yet exhibit phase separation into fcc- and bcc-based phases, as happens, for example, in Fe-Ni.\cite{BarabashFeX} We use two complementary approaches based on \emph{ab initio} calculations to determine the ordering tendencies within the given lattice type, focusing primarily on fcc alloys.

First, we apply the methodology of a cluster expansion (CE) with structural filters (SF), which was previously used to predict the ground states in the mixed-lattice Fe-(Ni,Pd,Pt) alloys.\cite{BarabashFeX,Chepulskii2012} This approach fully accounts for the atomic relaxations, including the changes in the shape of the unit cell. However, as explained below in Section \ref{CEmethod} and in Appendix \ref{appendix:CE}, the CE-SF approach to Au-Fe alloys meets with difficulties due to strong structure-dependent lattice instabilities. Therefore, we also employ an alternative methodology, the configuration-dependent lattice deformation model (CLDM),\cite{CLDM} which captures the effect of the local relaxations within the harmonic approximation, yet by construction excludes uniform strain and thus the possibility of a fcc-bcc transformation. CLDM can accurately describe the initial stages of ordering in second-order transitions, including ordering to \loz and \lott. We use CLDM to analyze the phase separation and ordering tendencies in fcc alloys, sorting out the contributions from the competing chemical and strain-induced interactions. The analysis of the long-range part of the strain-induced interaction is facilitated by the new ``simplified CLDM'' (S-CLDM) developed on top of the original CLDM.\cite{CLDM}

\subsection{{\em{Ab initio}} calculations}
\label{sec:DFT}

The {{\em{ab initio}} calculations have been performed within the generalized-gradient approximation (GGA-PBE)\cite{ref:PBE} to the density-functional theory (DFT).
We have employed the scalar-relativistic approximation with the pseudopotential projector-augmented wave method\cite{ref:PAW} (PAW) as implemented in VASP.\cite{ref:VASP}
Using the $T=0$ total energy of a structure $\sigma$ at Fe composition $x$, we calculate its zero-pressure formation enthalpy as
\beq
\Delta H(\sigma)=E_{tot}(\sigma)-x E_{tot}(\text{bcc Fe})-(1-x)E_{tot}(\text{fcc Au}).
\label{eq:DH}
\eeq

In principle, fcc Fe may exist in different low-spin (LS) and high-spin (HS) states; moreover, in pure fcc Fe, AFM-ordered and non-collinear spin configurations are energetically preferred\cite{NoncollinearFe} over the FM one. However, LS correlates with small atomic volume. Since the atomic volume of Au is larger, we expect the Au-Fe alloys to always be HS, except possibly at compositions very close to pure Fe. (Note that even pure fcc Fe, which is stabilized at high $T$, better correlates with disordered HS than with LS, and pure bcc Fe is always HS. Fe exhibits the LS state only when it is stabilized in the lower-lattice-parameter fcc structure.) Similarly, the preference for AFM ordering decreases with increasing lattice constant.\cite{NoncollinearFe} Indeed, we find that our test calculations converge to FM HS configurations, even if started with a LS initial magnetization (but with the volume near the expected HS value). We therefore limit our discussion to HS configurations. Special care has been taken to avoid numerical artifacts in the calculations of complex magnetic structures,\cite{Chepulskii2012} such as using small relaxation steps and an appropriate initial volume to avoid abrupt changes in the magnetic moments, and turning off VASP symmetrization in computationally problematic cases. Unless specified otherwise, the calculations have been performed for the FM state, as further justified in Sec.\ \ref{sec:magnetism}.

Structures used to construct CE-SF have been fully relaxed \cite{note:symrelaxation} using highly converged numerical settings; \cite{note:DFTdetails,ref:VOSKOWN} the Bain path calculations used similar settings at fixed geometry. For the CLDM construction, computational details were similar to Ref.\ \onlinecite{CLDM}. The CLDM input structures were first calculated with the ideal fcc positions and cell shape for the following lattice parameters: 3.810 \AA\ for 75\% Fe, 3.862 \AA\ for 66.7\% Fe, 3.953, 3.901, 3.8 and 3.7 \AA\ for 50\% Fe; the corresponding formation enthalpies (with respect to equilibrium fcc Au and bcc Fe) are referred to as $\Delta H_{chem}$.
The local (cell-internal) relaxations have then been allowed, while keeping the cell shape and volume fixed, resulting in $\Delta H_\text{fixed cell}$. Note that we used four different lattice parameters for the Au$_{0.5}$Fe$_{0.5}$ system in order to examine the volume dependence. The first value of 3.953 \AA\ for the Au$_{0.5}$Fe$_{0.5}$ system is the equilibrium lattice parameter of an undistorted 16-atom special quasi-random structure \cite{SQS} (SQS). The atomic volumes of this SQS and of pure Au and Fe were then fitted to a quadratic function, which was used to set the atomic volumes for AuFe$_{2}$ and AuFe$_{3}$ systems.
For Au$_{0.75}$Fe$_{0.25}$ we used $a=4.08$ \AA, which was obtained by minimizing the mean-squared volume relaxation energy for several input structures, as explained in Ref.\ \onlinecite{CLDM}. This value is very similar to the above-mentioned quadratic fit.

For the self-consistent calculations of the paramagnetic energy and Curie temperatures, we use the generalized-gradient approximation (GGA-PBE) and the coherent potential approximation (CPA) within the tight-binding linear muffin-tin orbital formalism in the atomic sphere approximation. The atoms are kept at the ideal fcc positions, and the equilibrium volume is used at each concentration. Full charge and CPA self-consistency are obtained for the total energy calculations. The paramagnetic state is represented by employing the disordered local moment (DLM) approximation, in which the Au$_{1-x}$Fe$_x$ alloy is represented by an auxiliary three-component Au$_{1-x}$Fe$^\uparrow_{x/2}$Fe$^\downarrow_{x/2}$ alloy, where Fe$^\uparrow$ and Fe$^\downarrow$ denote Fe atoms with local moments aligned parallel and antiparallel to the spin quantization axis.
The details of our implementation of CPA and DLM are described in Refs.\ \onlinecite{Ke,Belashchenko2015}. Equal sphere radii were used for Fe and Au; with this choice the sphere charges are approximately $0.1e$. We have checked that charge screening corrections for the Madelung potentials and total energy have a very small effect on the formation enthalpies and equilibrium lattice parameters. The numerical data reported below are without these corrections. From the difference in the formation enthalpies of the FM and paramagnetic (PM) states, the mean-field estimate of the Curie temperature is calculated as
\beq
T_C=(2/3)(\Delta H_\mathrm{PM}-\Delta H_\mathrm{FM})/x.
\eeq

\subsection{Cluster expansion with structural filters}
\label{CEmethod}

The CE method \cite{SanchezCE} maps the formation enthalpies (\ref{eq:DH}) of ordered structures onto an effective Ising-like Hamiltonian
\beq
\begin{array}{c}
\dH_{\text{CE}}(\sigma) =  J_0 +
 \sum_{f} J_{f} D_{f}\bar{\Pi}_{f}(\sigma).
\label{eq:CE}
\end{array}
\eeq
Here the actual geometrically relaxed configuration of Au and Fe atoms is mapped onto a configuration $\sigma$ of Ising pseudo-spins occupying the sites of an {\em ideal} parent (fcc or bcc) lattice, $f$ are the inequivalent geometric clusters of sites of the ideal lattice (such as pairs or three-body clusters of different size, etc., as well as the point cluster), the effective cluster interactions (ECIs) $J_0$ and $\{J_f\}$ are the coefficients of the generalized Ising Hamiltonian, while $D_{f}$ is the number per site and $\bar{\Pi}_{f}(\sigma)$ the correlation function in configuration $\sigma$ for cluster type $f$. Other physical quantities (e.g., the atomic volume) can be cluster-expanded instead of $\Delta H$ if desired. As long as the mapping between the relaxed geometries and the sites of the ideal lattice is unique, the infinite expansion (\ref{eq:CE}) is formally exact and unique.\cite{SanchezCE} It has been recently argued\cite{SanchezFourierCE} that, in general, the expansion (\ref{eq:CE}) does not formally converge. Nevertheless, the practical applications of a CE truncated to a finite number of terms have shown a surprising accuracy in predicting the DFT energies of new structures based on the parameters fitted to DFT energies of some ``input'' structures, particularly when using advanced techniques for selecting an ``optimal'' truncation for the expansion (\ref{eq:CE}).\cite{ATAT,CEwithGA,CEwithCS}

Our CEs are constructed using the ATAT package,\cite{ATAT} separately for fcc- and bcc-based structures, as further detailed in Appendix\ \ref{appendix:CE}. The values of the ECIs are fitted to the energies of an input set of structures $\sigma$ calculated in DFT. The energy of the structural relaxation is absorbed into the values of the ECIs.

The cornerstone of the CE methodology is the assumption that the relevant atomic configurations of the alloy (with the actual relaxed geometries) can be mapped one-to-one to the configurations of the Ising model defined on the underlying ideal lattice (such as fcc or bcc). However, this assumption is violated in Fe-Au alloys. The problems are two-fold: (1) Many structures are dynamically unstable. Some starting fcc configurations can relax all the way to nearly perfect bcc positions, and vice versa. This violates the uniqueness of the mapping: For example, even the simplest Bain-path transformation can be performed along different directions, resulting in three distinct mappings between fcc and bcc atomic positions.
(2) Different initial structures \emph{with the same lattice type} sometimes relax to the same structure. We will call such structures \emph{unmappable}. In a wide concentration range, the lowest-energy structures turn out to be unmappable. They are, in fact, ``hybrid'' superlattices (SL) with alternating layers of pure Fe and Au, which are close to their natural bcc and fcc geometries. For example, the (001), (011), and (111) bcc A$_3$B$_3$ SLs all relax to the same hybrid SL, which has the lowest DFT formation enthalpy among all structures with up to 6 atoms per unit cell.

Several methods have been suggested to extend the CE approach to mixed fcc/bcc alloys. One approach proposed by Liu {\em et al.}\cite{LiuFeCu} is to fix the cell shape and relax only the cell-internal coordinates. Another strategy, based on the concept of geometric filtering, was proposed in the earlier studies of Fe-(Ni,Pd,Pt) alloys.\cite{BarabashFeX,Chepulskii2012} For each structure $\sigma$, a ``score'' $s^{(\alpha)}(\sigma)$ of its proximity to the underlying lattice type $\alpha$ (fcc or bcc) is defined [see Eq.\ (\ref{eq:NNScore})], and the scaled ratio $r(\sigma)$ of the fcc and bcc scores [Eq.\ (\ref{eq:NNScoreR})] is used as a structural filter (SF) to classify the structure as fcc-like or bcc-like.

Here we follow the CE-SF prescription, constructing separate fcc and bcc CEs for the FM HS Au-Fe alloys, each including only structures that retain the given lattice type after relaxation. However, due to the problem (2) mentioned two paragraphs above, the structural filtering alone is not sufficient to make the CEs meaningful for Au-Fe alloys, and we have also excluded all unmappable structures from the input sets. Further details are included in Appendix\ \ref{appendix:CE}.

\subsection{CLDM and S-CLDM}
\label{CLDMmethod}

The ordering tendencies in a (nearly) random alloy at constant pressure can be considered as the coefficients of the second-order expansion of the Gibbs free energy with respect to small deviations from homogeneity. In view of the large size mismatch in Au-Fe alloys, it is imperative to include the contribution of structural relaxations. The displacements of atoms under structural relaxation can be represented as a superposition of macroscopic strain (change in volume and shape of the unit cell) and local displacements. While local displacements contribute to the second-order expansion of the Gibbs free energy, homogeneous strain does not, as explained in Appendix \ref{app:proof}. Therefore, in the study of the ordering tendencies we need to consider only local relaxations induced by ordering, while keeping the macroscopic strain (cell shape and volume) fixed.
In other words, the energetics of disordered (and the approximate energetics of weakly-ordered) alloys is given by
\beq
\Delta H_\text{fixed cell}(\sigma) = \Delta H_{chem}(\sigma) + E_{rel}(\sigma),
\label{eq:DHchem}
\eeq
where $\Delta H_{chem}$ represents the ``chemical'' formation enthalpy computed with all atoms kept at ideal fcc positions, and $E_{rel}$ is the energy gained by local atomic relaxations at {\em{constant}} uniform strain, i.e., for periodic structures, at fixed shape and volume of the unit cell. We construct a CE (referred to as ``chem-CE'') for the chemical term $\Delta H_{chem}$, which depends only on the local environment. However, we want to avoid cluster-expanding $E_{rel}$, because strain-induced interaction is long-ranged and singular at large distances.

To properly describe the strain-induced interaction, we employ the configuration-dependent lattice deformation model (CLDM),\cite{CLDM} which generalizes the Kanzaki-Krivoglaz-Khachaturyan model \cite{Matsubara,Kanzaki,Krivoglaz,Khachaturyan} to the case of a concentrated alloy. The many-body, long-range strain-induced interaction is described in the harmonic approximation by the relaxation energy
\begin{equation}
E_{rel}(\sigma)=-\frac12\sum_{ij}\mathbf{F}_i(\sigma) \hat A^{-1}_{ij}(\sigma)\mathbf{F}_j(\sigma),
\label{erel}
\end{equation}
where $\mathbf{F}(\sigma)$ and $\hat A(\sigma)$ are the configuration-dependent Kanzaki forces and force constants, and the summation is over the lattice sites.

The CLDM is constructed for a fixed concentration and describes the relaxation energy under the assumption that the crystal lattice remains fully coherent. The parametric dependence of the effective Hamiltonian on the average composition is a general feature of coherent phase transformations.\cite{CLDM}

Both $\mathbf{F}(\sigma)$ and $\hat A(\sigma)$ are represented by separate many-body cluster expansions. Note that, even though both these CEs are short-range, the inversion of the force constant matrix leads to a long-range expression (\ref{erel}), properly capturing this feature of the strain-induced interaction. $\mathbf{F}(\sigma)$ is fitted directly to the results of DFT calculations for interatomic forces in structure $\sigma$ at the ideal fcc positions. Specifically, the force acting at site $i$ is taken to depend on the identity (and the relative positions) of the atom at site $i$ and some of its neighbors, as further detailed in Appendix\ \ref{app:CLDM}.

The force constants are determined using linear regression for the set of equations
\beq
\delta {\bf F}(\sigma,\mathbf{u})=\hat A(\sigma) \mathbf{u},
\eeq
where $\delta\mathbf{F}(\sigma,\mathbf{u})$ are the changes in the DFT forces arising due to small atomic displacements $\mathbf{u}$. The sample set of $\delta \mathbf{F}(\sigma,\mathbf{u})$ was calculated using the VASP code, as described in Ref.\ \onlinecite{CLDM}. For the force constants we used a simple parametrization, in which only central (bond-stretching) interactions depend on the configuration, while the non-central interactions are configuration-independent.

We construct a separate CLDM for each given composition and lattice parameter. We have considered compositions of 25, 50, 66.7 and 75\% Fe. In addition to the lattice parameters designed to represent equilibrium volumes, several additional lattice parameter values have been taken for the Fe$_{0.5}$Au$_{0.5}$ system in order to examine the volume dependence, as discussed above in Sec.\ \ref{sec:DFT}.
In order to reduce systematic errors in chem-CE, the values of $\Delta H_{chem}$ have been calculated for the same set of input structures (covering all of the above compositions) at each lattice parameter. We have later discovered that CLDM predicts random alloys at 25\% Fe to be dynamically unstable, which makes CLDM approach inapplicable at that composition. The parameters of the cluster expansions for the Kanzaki forces and force constants at other compositions, as well as the details of chem-CE construction, are presented in Appendix\ \ref{app:CLDM}.

Ordering tendencies in nearly-random alloys can be characterized by considering an ensemble in which the average occupation $\bar\sigma_i$ at site $i$ differs only slightly from the average over all sites $\sigma_0$, i.e., $\bar\sigma_i = \sigma_0 +\delta_i$, where all $\delta_i$ are small. The effective pairwise interaction potential is then defined as the second derivative of the ensemble average of the energy:
\beq
J^\mathrm{eff}_{ij} =\frac{\partial^2 \langle E\rangle}{\partial\delta_i \partial\delta_j}.
\label{eq:Jderiv}
\eeq
Its Fourier transform gives $J_\mathrm{eff}({\bf k})$, which within CLDM can be readily decomposed into chemical and strain-induced contributions, representing the respective terms in Eq.\ (\ref{eq:DHchem}):
\beq
J_\mathrm{eff}({\bf k})=J_{chem}({\bf k})+J_\mathrm{SI}({\bf k}).
\label{eq:Jchem}
\eeq
Note that $J_{chem}$ reflects purely chemical trends, even though the ``chemical'' term $\Delta H_{chem}$ in Eq.\ (\ref{eq:DHchem}) includes the ``volume deformation energy,'' \cite{LaksMBCE} which is the elastic energy required to bring the atoms of the constituent elements to the common lattice parameter, prior to any further relaxation. This is because the volume deformation energy is configuration-independent at the given composition. Similarly, $J_\mathrm{SI}$ captures all strain-induced interactions pertaining to deviations from the random alloy.

In order to compute the strain-induced term $J_\mathrm{SI}(\mathbf{k})$ from CLDM, Ref.\ \onlinecite{CLDM} employed an additional fitting of $E_{rel}$ computed from Eq.\ (\ref{erel}) for a few hundred structures to a multiparametric real-space many-body CE. This CE for $E_{rel}$ was added to the chem-CE, whereupon the second derivative in Eq.\ (\ref{eq:Jderiv}) leads to Eq.\ (\ref{jeff}) in Appendix \ref{app:CLDM}. Although this procedure allows one to retain a large number of terms in the CE, it still misses the true long-range character of the strain-induced interaction and its singularity at the $\Gamma$-point. To remedy this deficiency, we have developed another method of extracting $J_\mathrm{eff}(\mathbf{k})$ from CLDM. The idea is to find a simplified form of CLDM (S-CLDM), with configuration-\emph{independent} force constants, that would approximately reproduce the full $E_{rel}$ predicted by CLDM, while also allowing a simple calculation of $J_\mathrm{eff}(\mathbf{k})$ without the additional CE expansion. Although the existence of such a simplified form is not guaranteed \emph{a priori}, we have found that in Au-Fe alloys it can be constructed. This S-CLDM captures the dominant part of the full CLDM, while the remainder of $E_{rel}$, only a few meV/atom in magnitude, can be fitted to a separate ``residual'' CE. The details are given in Appendix\ \ref{app:CLDM}.

\section{FCC-BCC lattice stability}
\label{sec:fccbcc}

In this section we analyze the general energetics of fcc and bcc alloys, without focusing on the specific identity of the ordered phases. Our purpose is to determine whether the {\em bulk} ordering and phase competition tendencies, combined with the single assumption of the {\em suppression of phase separation and spinodal decomposition} are  sufficient to explain the experimental observations of (a) the bcc phase in as-deposited nanoparticles with 65\% Fe or more and fcc phase with 53\% Fe or less,\cite{Shield2013} and (b) transformation to the fcc phase after annealing of the nanoparticles with 79\% Fe.\cite{Shield2014} Further, we analyze (c) whether the stability of fcc \emph{vs} bcc lattice type is determined primarily by temperature and concentration, regardless of the atomic configuration and magnetic ordering. Finally, while not aiming to settle whether such a suppression of phase separation could be of a thermodynamic or a kinetic origin, we would like to limit its possible nature, asking (d) whether inhibiting the decomposition into dissimilar bcc and fcc phases is sufficient to observe miscibility and formation of ordered phases. For example, Fe-Ni and Fe-Pd alloys have been shown\cite{BarabashFeX,Chepulskii2012} to exhibit a strong tendency to form ordered compounds if restricted to the fcc lattice, and the wide miscibility gap seen in these alloys at Fe-rich compositions is solely due to the freedom to precipitate out the bcc phase. If the Au-Fe alloys exhibit similar energetics, then inhibiting the separation into fcc and bcc phases (for example, due to high interface energy penalty) could be sufficient to stabilize ordered phases in Au-Fe nanoparticles; otherwise, there is an inherent tendency for a compositional disproportionation even {\em within} the same (all-fcc or all-bcc) lattice system.

\subsection{Miscibility and fcc/bcc competition}
\label{sec:fccbcc:random}

Fig.\ \ref{fig:DeltaHunconstrained} shows the formation enthalpies $\dH$ of fcc-based (black) and bcc-based (red) structures at $T=0$, calculated in DFT and fitted to the CE-SF. Clearly, there is a thermodynamic driving force toward phase separation, even disregarding the competition between fcc and bcc lattices. Indeed, the energies of all periodic structures are larger (by 51 meV/atom or more at $x=0.5$) than the average of pure Fe and Au energies for the \emph{same} lattice type. Within the classical Miedema model, \cite{Miedema} this tendency toward phase separation originates from the relatively small charge-transfer energy gain, implied by the small difference in the work functions (or Allen electronegativities) of Fe and Au, which is too small to overcome the electron density mismatch. This argument applies separately to fcc and bcc alloys. Moreover, the Miedema model disregards the positive contribution to $\dH$ from the elastic strain due to the large size mismatch between Fe and Au, which should further increase the miscibility gap. Thus, both chemical and elastic terms favor phase separation in bulk fcc Au-Fe alloys, even if the precipitation of the bcc phase is inhibited; this is in contrast to the ordering tendency exhibited by fcc-restricted Fe-Ni and Fe-Pd alloys.

\begin{figure}[hbt]
\begin{center}
\includegraphics[width=0.45\textwidth,clip]{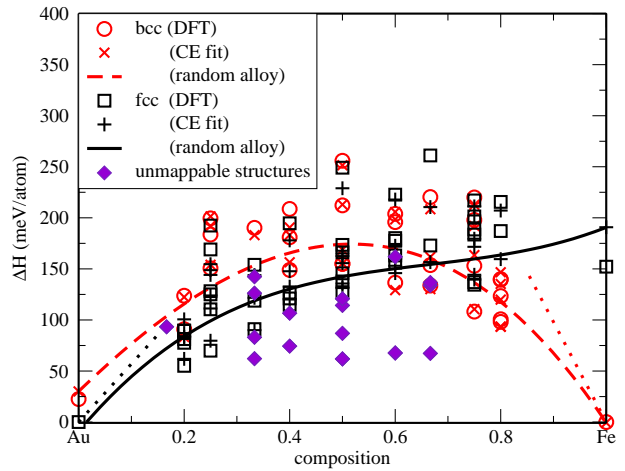}
\end{center}
\caption{(Color online) Formation enthalpies of ferromagnetic, fully relaxed fcc (black) and bcc (red) ordered Au-Fe structures (only those with up to 5 atoms per cell are shown). Open symbols: DFT results; crosses or pluses: CE predictions; lines: CE predictions for random alloys (see legend). Diamonds (purple): unmappable structures (see text). Dotted lines: asymptotic tangents estimated for dilute alloys (see Appendix E).}
\label{fig:DeltaHunconstrained}
\end{figure}

When the experimental deposition is performed at low temperature (as in Ref.\ \onlinecite{Shield}), the kinetic barrier to atomic ordering and spinodal decomposition is high, and the Fe and Au atoms within the nanoparticle structure likely stay disordered, nearly random. The solid (black) and dashed (red) lines in Fig.\ \ref{fig:DeltaHunconstrained} show the $T=0$ formation enthalpies of the random fcc and bcc alloys predicted by the cluster expansions. The concave-down curves do not directly indicate spinodal instability of the random alloys, because coherent spinodal decomposition may be blocked by the coherency strain energy (see further discussion in Sec.\ \ref{sec:disordered:spinodal}). Note that the CE predictions for dilute alloys should be treated as extrapolations. Further discussion can be found in Appendix \ref{app:defects}, where the enthalpies of dissolution of Fe in fcc Au and of Au in bcc Fe are calculated using large supercells. In both cases, the corresponding slopes of the formation enthalpies, which are shown by the dotted lines in Fig.\ \ref{fig:DeltaHunconstrained}, agree reasonably well with the random-alloy lines predicted by the CEs.

The crossing of the solid and dashed lines in Fig.\ \ref{fig:DeltaHunconstrained} indicates that, at low $T$, the random bcc alloy is preferred over fcc if the concentration of Fe exceeds 68\%. This is a relatively crude estimate due to the finite CE accuracy (see Table\ \ref{table:CEparams} in Appendix\ \ref{appendix:CE}). For example, a different fcc CE with the same set of inputs and a nearly identical cross-validation score, but with a larger number of three-body ECIs, moves the fcc/bcc crossing to 63\% Fe. Experimentally,  as-deposited Au-Fe nanoparticles (prepared by inert gas condensation at 143 K) are observed in the fcc phase at and below 53\% Fe, and in the bcc phase at and above 65\% Fe,\cite{Shield2013} which agrees with the CE predictions qualitatively and, within the CE accuracy, quantitatively.

The observed bcc lattice parameter was anomalously large, which could be due to the large concentration of vacancies.\cite{Shield2013}  Here we disregard the effect of vacancies both on the lattice parameter and on the phase stability, even though they may modify the stability range of the as-deposited bcc phase. In our calculations the lattice parameter of the bcc structures is close to the Vegard law predictions, much smaller compared to the experimental values.

Upon heating, nanoparticles may undergo a diffusionless martensitic transformation from the bcc to fcc phase. The martensitic temperature $T_\text{mart}$ may be defined as one where the Gibbs free energies of random single-phase fcc and bcc alloys are equal. We do not attempt to calculate $T_\text{mart}$ from first principles, because the vibrational and magnetic contributions are very difficult to capture accurately in Fe alloys.\cite{Leonov}
However, a crude estimate can be made by combining our results at $T=0$ with the data from the experimental phase diagram. This is illustrated in Fig.\ \ref{fig:phD}, where we redraw the experimental phase diagram based on data from Ref.\ \onlinecite{ASM}. Point A (red dot) marks the concentration $x_0$ where $T_\text{mart}(x_0)=0$, according to the CE estimate illustrated in Fig.\ \ref{fig:DeltaHunconstrained}. At Fe-rich compositions $T_\text{mart}(x)$ must lie within the two-phase $\alpha+\gamma$ region CDE. Therefore, assuming that the curvature $d^2T_\text{mart}(x)/dx^2$ does not change sign as a function of $x$, the $T_\text{mart}(x)$ curve must lie somewhere within the hashed (red) region ABCDEA in Fig.\ \ref{fig:phD}. Clearly, at the annealing temperature of 873 K the fcc random alloy is predicted to be stable at all concentrations (33-79\% Fe) studied in Refs.\ \cite{Shield,Shield2014}, in agreement with the observation of the fcc structure in all annealed nanoparticles.

\begin{figure}[bt]
\begin{center}
\includegraphics[width=0.45\textwidth]{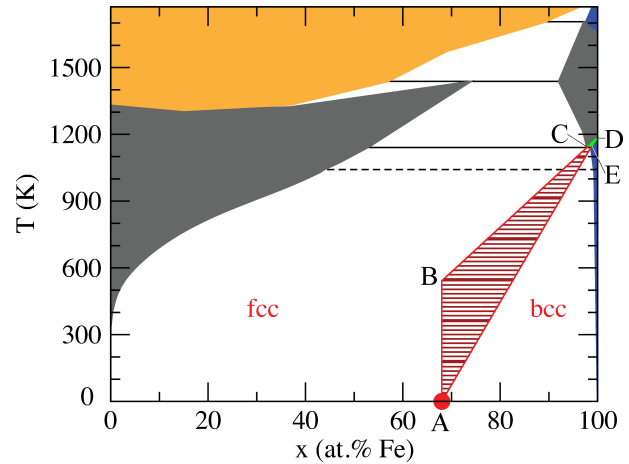}
\end{center}
\caption{(Color online) Phase diagram of bulk Au-Fe based on the experimental data from Ref. \onlinecite{ASM}. Shaded areas: single-phase regions (blue: bcc; gray: fcc; yellow: liquid). Point A: predicted intersection of the bcc and fcc random-alloy formation enthalpies from Fig.\ 1. Point C: eutectoid point. CDE (green): two-phase ($\alpha+\gamma$) region. Point B lies directly above point A on the continuation of the DC line. The martensitic transformation line $T_\text{mart}(x)$, which shows equilibrium between random single-phase fcc and bcc alloys, is predicted to lie inside the hashed (red) area ABCDEA. Labels ("fcc" and "bcc") indicate the regions of relative stability of these random alloys.}
\label{fig:phD}
\end{figure}

\subsection{Configurational dependence of fcc/bcc transformations}
\label{sec:fccbcc:ordered}

We found that about one third of the structures initially at fcc positions attain lower energy by relaxing towards bcc. Surprisingly, different ordered bcc structures at the {\em same} composition may exhibit an opposite transformation towards fcc. This is in contrast to what was found for the Fe-Pd system, \cite{Chepulskii2012} where the fcc-bcc instabilities of ordered alloy structures strongly correlate with the alloy composition.

As mentioned above, in some cases two or more initial structures of the same lattice type relax toward the same final structure. These unmappable structures are shown in Fig.\ \ref{fig:DeltaHunconstrained} by diamonds, and they include hybrid bcc/fcc SLs in a wide range of concentrations (33-67\% Fe). Such SLs have the lowest formation enthalpy, reflecting the tendency to phase separation even if the lattice remains coherently strained.

In Fig.\ \ref{fig:ScoreVsX}, we illustrate the proximity of final atomic positions \cite{note:relaxation} of different fully relaxed structures to the fcc and bcc geometries, as measured by the scores defined by Eqs.\ (\ref{eq:NNScore}) and (\ref{eq:NNScoreR}). We see that at nearly all compositions there are both bcc-unstable and fcc-unstable structures.  The instabilities of ordered structures in Au-Fe seem to correlate with the direction of ordering vectors. For example, we found that fcc (110) SLs, i.e., structures composed of (110)-oriented atomic planes of pure Fe and pure Au, all relax to bcc, whether they are AuFe$_3$, Au$_2$Fe$_2$, or Au$_3$Fe SLs, whereas (001), (201) or (311) SLs retain the original fcc lattice at all these compositions. The typical fcc$\rightarrow$bcc transformation route during DFT relaxation, which was common to all (110) SLs, is a tetragonal collapse along the [001] direction. In this regard, the stability of the tetragonal (001) SLs appears surprising. In a few cases the relaxation history has been much more complex: for example, one Au$_3$Fe$_2$ structure
has switched several times between fcc-like and bcc-like geometries. One of the structures had two local minima, both belonging to the fcc lattice type.

\begin{figure*}[th]
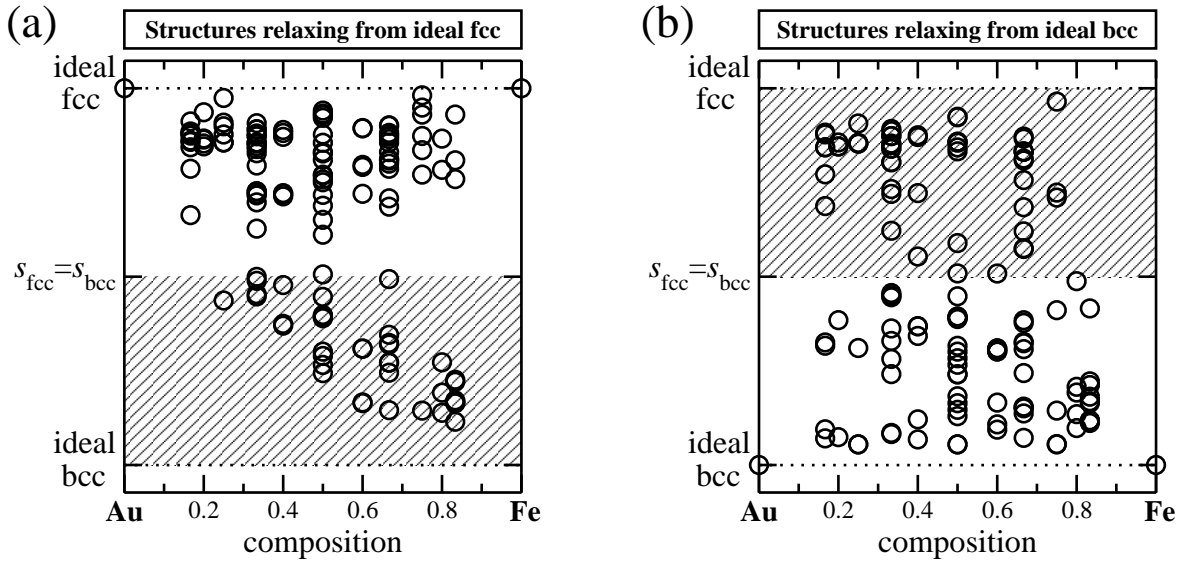

\begin{center}
\vskip1pc
\includegraphics[width=0.4\textwidth,clip]{Fig3a.eps}
\hfil
\includegraphics[width=0.4\textwidth,clip]{Fig3b.eps}
\end{center}
\caption{Final scaled ratio of fcc/bcc scores after geometry relaxation for structures initially at (a) ideal fcc and (b) ideal bcc positions. The shaded areas highlight structures that relax to a different type of lattice geometry.}
\label{fig:ScoreVsX}
\end{figure*}

While our CEs predict the formation enthalpies of the random alloys, they give no information about the dynamical stability with respect to the fcc-bcc transformation. DFT calculations for SQSs suggest that the fcc random alloy may become dynamically unstable with respect to a martensitic transformation at high Fe content and low temperatures. For example, at Au$_{1/3}$Fe$_{2/3}$ composition both fcc-based SQSs that we have considered relaxed without barrier towards bcc-like geometries. (In contrast, both bcc-based SQSs remained bcc, and their formation enthalpies agree to 11 and 23 meV/atom with the CE prediction for the random bcc alloy.)
Dynamical instability of the random fcc alloy would imply that {\em random} nanoparticles, even if stabilized in fcc during the anneal, would tend to transform back to the bcc geometry upon cooling. Based on these results, it could be argued that some degree of order has to develop during the anneal for the fcc geometry to remain stable against martensitic transformation to bcc.

However, we do not find any dynamical instabilities in CLDM simulations at the Au$_{1/3}$Fe$_{2/3}$ composition. The likely reason for this apparent contradiction is that an fcc SQS does not represent the behavior of the random alloy along the fcc-bcc transformation path, because the nearest-neighbor shells, for which the quasi-random correlations only hold, change during the transformation. As a result, the SQS structures lose their ``quasi-randomness'' after the martensitic transformation. The deficiency of SQS in representing the fcc-bcc transformation is apparent from the fact that the formation energies of the fcc SQSs that have relaxed to bcc (98 and 96 meV/atom) are much lower compared to the random bcc alloy predicted by the CE (160 meV/atom). On the other hand, the absence of unstable phonon modes predicted by CLDM does not guarantee that the fcc structure is stable with respect to a homogeneous strain deformation.

\subsection{Magnetic effects and fcc-bcc transformation path}
\label{sec:magnetismBain}

It is possible that the temperatures at which the nanoparticles are annealed \cite{Shield,Shield2014} could be above their Curie temperatures. Furthermore, it was suggested \cite{Shield2014} that \loz AuFe and \lot Au$_3$Fe could favor AFM ordering. As in the case of pure Fe,\cite{Singh1991} the stability of the Au-Fe alloy with respect to the fcc-bcc transformation could depend strongly on the magnetic contribution. Therefore, here we consider the effect of magnetic ordering on the fcc-bcc transformation. Further analysis of the effects of magnetism on the ordering tendencies is postponed till Sec.\ \ref{sec:magnetism}.

We limit ourselves  to the case of the AuFe \loz ordered structure, which has been reported most often in Au-Fe.\cite{Takanashi,Sato,Shield,Shield2014}
We consider the change in the energy of \loz along the fcc-bcc transformation path for several types of magnetic order, including an approximate model for the PM state. In order to approximate the energy of the PM \loz phase, we average the energies of different spin orderings. These energies are calculated at the atomic positions maintaining the symmetry of the PM state, under the assumption that spin fluctuations occur on a shorter time scale compared to ionic displacements, as in the Born-Oppenheimer approximation for the electronic degrees of freedom. Conveniently, in the case of \loz ordering such ``average'' atomic positions are fully determined by the total volume $V$ and the $c/a$ ratio. In turn, the equilibrium $V$ and $c/a$ values are determined by the condition of the vanishing diagonal stress components $\sigma_{xx}=\sigma_{yy}$ and $\sigma_{zz}$.
We then analyzed how the total energy $E$ and $\sigma_{xx}$, $\sigma_{zz}$ depend on the geometry in different spin states. The dependence on $V$ was found to be insignificant, but the variation with $c/a$ is of great interest. The results presented below were obtained for the volume fixed at its equilibrium value for the FM state.

It is well known that by changing $c/a$ one may convert an fcc geometry ($c/a=1$) into a bcc geometry ($c/a=1/\sqrt{2}$) along the so-called Bain path.\cite{note:Bain} This is indeed the mode of collapse of many fcc-unstable structures in our DFT calculations. Plotted along the Bain path, the energy of pure Au has a global minimum at fcc, a {\em{maximum at bcc}}, and a secondary bct minimum at a slightly smaller $c/a$ value, \cite{MehlBain2004,SchoneckerThesis} whereas the energy of pure Fe, in the FM HS state at $T=0$, has a global minimum at bcc,  a {\em{maximum at fcc}}, and a secondary minimum at a slightly larger $c/a$. (The latter minimum is sometimes referred to as ``fct'' geometry, to stress its proximity to fcc, despite its crystallographic equivalence to bct). It has been demonstrated \cite{MehlBain2004} that the secondary minima in the Bain paths of most elemental metals are, in fact, saddle points, which are unstable under orthorhombic distortions. This instability is not generic for the bct and fct geometries and only reflects the energetics of the common pure elements. Indeed, the bct geometry was argued to correspond to the structure of Pa and of $\beta$-Hg, whereas In is observed in a structure similar to the fct geometry.\cite{MehlBain2004} Thus, it is natural to expect that Au-Fe alloys may similarly have two energy minima along the Bain path, and that the higher-energy minimum may be dynamically unstable for some structures and stable for others.

\begin{figure*}[hbt]
\begin{center}
\includegraphics[width=0.9\textwidth]{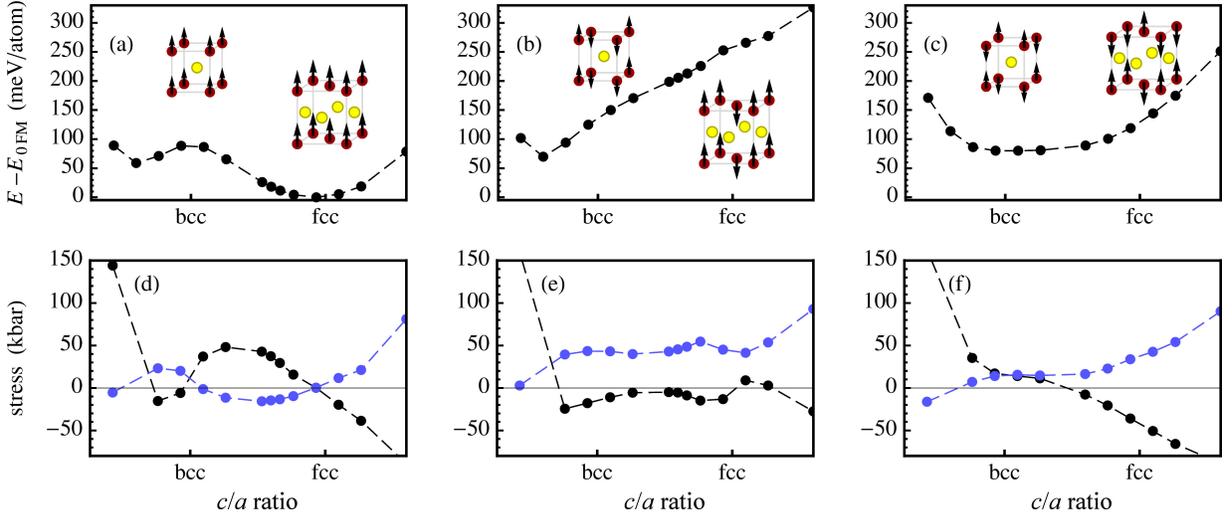}
\end{center}
\caption{(Color online) Total energy (a-c) and stress tensor components (d-f) as a function of the $c/a$ ratio plotted along the volume-conserving Bain path of \loz AuFe with FM (a,d), C-type AFM (b,e) and G-type AFM (c,f) spin orderings. The volume of the equilibrium FM phase is used throughout, and it is also taken as the energy reference. The stress components are $\sigma_{xx}=\sigma_{yy}$ (light blue dots) and $\sigma_{zz}$ (black dots). Insets in panels (a-c) show structural and magnetic order at the bcc and fcc values of the $c/a$ ratios. Dashed lines are guides for the eye.
}
\label{fig:Bain}
\end{figure*}

Fig.\ \ref{fig:Bain} shows $E$ and $\sigma_{xx}$, $\sigma_{zz}$ along the Bain path of \loz AuFe, at a constant volume, for three different spin orderings: the FM state and two AFM orderings (C-type and G-type) illustrated in the insets in Fig.\ \ref{fig:Bain}(a-c). The ordering vectors are $(1,0,0)$ for the C-type and $(1,0,1/2)$ for the G-type state. These vectors are given in units of $2\pi/a$ (or $2\pi/c$ for the $z$ component); the notation is similar to Ref.\ \onlinecite{Pujari}.

The Bain energy profile of the FM \loz phase [Fig.\ \ref{fig:Bain}(a)] is similar to that of Au,\cite{MehlBain2004,SchoneckerThesis} with a global minimum close to the ideal fcc $c/a$ ratio. Surprisingly, the addition of bcc-stable Fe has raised the relative energy of the bct minimum instead of lowering it; in pure Au, bct is only 20 meV/atom above the global fcc minimum.\cite{MehlBain2004}) However, the Bain-path profiles for both AFM phases show a {\em single} deep minimum near the bcc value of $c/a$ and no minimum near fcc. The absence of a local fcc-like minimum in the AFM-ordered \loz phases is particularly clear from the plots of $\sigma_{xx}$ and $\sigma_{zz}$, which intersect at a single point, near the bcc $c/a$, for both AFM orderings. Due to the relatively small elastic anisotropy, the volume changes shift the plots of $\sigma_{xx}$ and $\sigma_{zz}$ almost rigidly up or down, and thus do not help achieving $\sigma_{xx}=\sigma_{zz}=0$ around the fcc $c/a$. We conclude that both AFM orderings of the \loz phase shown in Fig.\ \ref{fig:Bain} are dynamically unstable in the fcc-like geometry.

Although both C-type and G-type orderings have antiparallel nearest neighbors, Fig.\ \ref{fig:Bain} shows that their energies are very different. There are two reasons for that. First, the magnetic interactions from more distant neighbors contribute substantially to the AFM energies (which is not the case in the FM and PM cases, see Sec.\ \ref{sec:magnetism}). We explicitly fitted the magnetic interactions to a generalized Ising model at the fcc positions and found that the FM coupling to third-nearest and AFM coupling to fourth-nearest fcc neighbors are both $\sim 15$\% of the nearest-neighbor coupling, while involving twice as many Fe atoms. While these opposite-sign contributions largely cancel out in both FM and PM states, they add up in the AFM states, raising and lowering the energy of the C-type and G-type orderings, respectively. Second, as the $c/a$ ratio is decreased toward its bcc value, the tetragonal \loz structure becomes cubic B2 with the Fe sites forming a simple cubic sublattice. In this structure, all nearest-neighbor pairs are antiparallel if the ordering is G-type, but in the C-type ordering only four of the six nearest-neighbor pairs are antiparallel. This has an important symmetry implication: the FM and G-type orderings have full cubic symmetry at the bcc value of the $c/a$ ratio, and, therefore, they must have either a maximum (the FM case) or a minimum \cite{note:AFMBainSymmetry} (the G-type AFM case) at the bcc positions. In contrast, the symmetry of the C-type AFM phase remains tetragonal, and the minimum is achieved at a smaller $c/a$ ratio.

The energy of the PM state can be approximated by averaging the energies of appropriately chosen magnetically ordered states. Any reasonable spin averaging should respect the cubic symmetry of the PM phase at bcc positions. As mentioned above, the FM and G-type orderings already respect this symmetry, and the simplest approximation for the PM energy can be obtained by taking their average: $E_\mathrm{PM} \approx (E_\mathrm{FM}+E_\mathrm{G})/2$. This estimate (Model 1) is displayed by a gray line in Fig.\ \ref{fig:BainPM}. Model 1 averages out any linear function of nearest-neighbor spin correlators, which means it should give the correct PM energy if the exchange interaction is dominated by nearest-neighbor Heisenberg exchange.

An alternative estimate for the PM energy can be obtained by including C-type ordering in the average. Since it is tetragonal, it should be included along with all of its images obtained by applying cubic symmetry operations at the bcc positions. The rotation around one or the other of the two in-plane cubic axes of the bcc structure produces equivalent spin orderings C$_\pm$ characterized by the ordering vectors (1/2,$\pm1/2$,1/2). At any $c/a$ ratio, the C$_\pm$ structures are related by a 90$^\circ$ rotation around the $z$ axis, and their Bain paths (not shown in Fig.\ 4) are identical with a minimum near fcc but no minimum near bcc. A straightforward enumeration shows that the estimate $E_\mathrm{PM} \approx (E_\mathrm{FM} + E_\mathrm{C} + 2 E_\mathrm{C_\pm})/4$ averages out any linear function of pairwise spin correlators for four nearest coordination spheres along the entire Bain path. This estimate (Model 2) is shown by the black line in Fig.\ \ref{fig:BainPM}.

Models 1 and 2 give somewhat different estimates of the PM energy but agree in their main features: the PM energy profile along the Bain path is qualitatively similar to the FM state, having a maximum at bcc and two minima with $c/a$ somewhat below the fcc and bcc values, of which the global minimum corresponds to the fcc-like geometry. On the other hand, magnetic disorder substantially decreases the $c/a$ ratio and increases the energy of the fcc-like minimum, while having a smaller effect on the bcc-like minimum. As a result, both the energy difference and the barrier separating the two minima along the Bain path decrease. Note that this trend is opposite to what is observed in pure Fe, where ferromagnetism stabilizes the bcc phase relative to fcc. The reason is that bcc Fe and \loz AuFe are both ferromagnetic, whereas the fcc Fe is magnetically frustrated, and its energy is, therefore, less sensitive to magnetic ordering.\cite{Moruzzi,NoncollinearFe}

To conclude this section, random deviations from FM order may decrease the bcc-fcc energy difference and the $c/a$ ratio of the fcc-like L1$_0$ phase, but otherwise they do not qualitatively change the fcc-bcc transformation path. This does not mean that the magnetostructural coupling is intrinsically weak, because, as seen in Fig.\ \ref{fig:Bain}, enforcing AFM order with antiparallel nearest neighbors drastically changes the fcc-bcc energetics.

\begin{figure}[hbt]
\begin{center}
\includegraphics[width=0.4\textwidth]{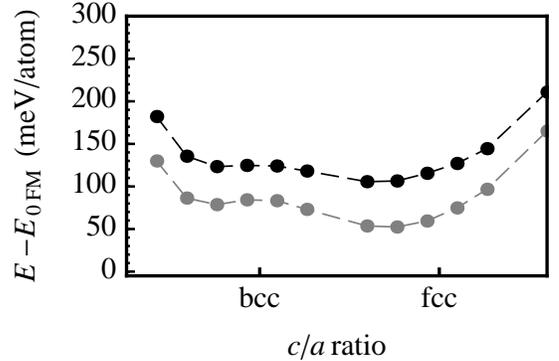}
\end{center}
\caption{Energy as a function of the $c/a$ ratio, along the volume-conserving Bain path, for the paramagnetic \loz\ AuFe. Models 1 (gray line) and 2 (black line) average over different magnetic states (see text). The volume of the equilibrium FM phase is used throughout, and it is also taken as the energy reference.
}
\label{fig:BainPM}
\end{figure}

\section{Energetics of fully ordered alloys}
\label{sec:ordered}

\subsection{Fully relaxed structures}
\label{sec:ordered:CE}

In this section we analyze the energetics of fully ordered structures using the CE-SF approach, imposing no restrictions on the geometric relaxation. Table \ref{table:GSdH} lists the formation enthalpies for several structures, to be discussed below, in the first two \mdH columns.

Before proceeding, we emphasize the key drawback of the CE-SF method. Despite the fcc-bcc filtering of the \emph{input} structures, the \emph{predictions} of, say, an fcc CE-SF include not only fcc-stable structures, but also the hypothetical structures that are, in reality, fcc-unstable at $T=0$. Such predictions could perhaps give reasonable approximations for the fcc energy of an ordered region under some fcc-stabilizing conditions, such as elevated temperature or coherency strain, but they should be treated as unreliable extrapolations. Moreover, the CE-SF does not tell us whether a given structure is fcc-stable; this needs to be checked by a DFT calculation. However, if the lowest-energy structure predicted by the fcc CE-SF is fcc-unstable, then there must be an infinite number of fcc-unstable structures below the lowest-energy fcc-stable structure. In this case it becomes impossible to unambiguously identify the lowest-energy fully relaxed fcc-stable structure. In such cases a CE constructed at the fixed cell shape may be preferable.\cite{LiuFeCu} If the assumed structure is stabilized by the phonon entropy, more accurate estimates of the fcc-stabilized energies could also be obtained using \emph{ab initio} molecular dynamics.\cite{OzolinsAIMD}

\begin{figure}[!h]
\begin{center}
\includegraphics[width=0.45\textwidth]{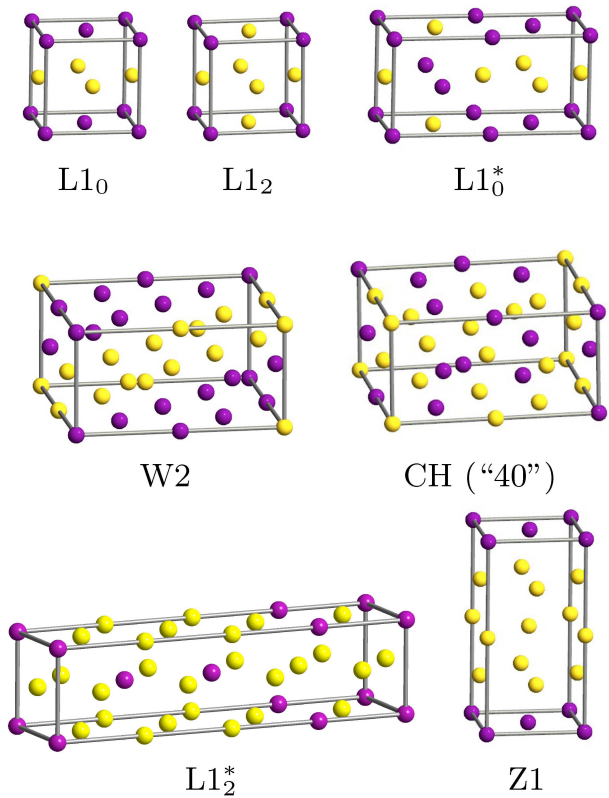}
\end{center}
\caption{(Color online)
The low-energy ordered Au-Fe structures, as determined by different approaches (cf. Table\ \ref{table:GSdH}).
The conventional cells are shown in most cases, except  W2 is displayed within a non-periodic $2\times2\times1$ fcc unit for clarity.
}
\label{fig:structures}
\end{figure}

\begin{table*}[!ht]
\caption{Formation enthalpies of the low-energy ordered Au-Fe structures as determined by different approaches. The fcc structures listed here are illustrated in Fig.\ \ref{fig:structures}.
The lattice parameters of the ``fixed cell'' and ``unrelaxed'' structures are chosen as discussed in Sec.\ \ref{sec:DFT}. The S-CLDM+ column includes the contribution from the residual CE, and S-CLDM does not (see Appendix \ref{s-cldm} for details).
}
\begin{tabular}{c|c c| c | c | c | c | c | c | c | c | c}

\hline
\hline
Composition  	& Structure & Lattice & \multicolumn{8}{c|}{Formation enthalpy \mdH (meV/atom) } 							& Reason \\
\cline{4-11}
			&		& type & \multicolumn{2}{c|}{Fully relaxed} & \multicolumn{4}{c|}{Fixed cell
													($\Delta H_\text{fixed cell}$)}  &\multicolumn{2}{c|}{Unrelaxed
																				($\Delta H_{chem}$) } & to expect \\
\cline{4-11}
			&		&	 & GGA		&  CE-SF		&	GGA &  CLDM	& S-CLDM  & S-CLDM+	&  GGA\hfil	& CE	& lowest \mdH	\\		
\hline

Au$_3$Fe 	& \lot 		&  fcc 	& 192.4		& 150.7		& 193.9	& 191.6	& 191.6	& 189.2	&  193.9	& 191.6			& Experiment\\
			& Z1			& fcc	& 69.9		& 79.6		&  70.8	& 69.2	&  87.8		& 69.7 &  130.6	& 132.2			& CE-SF \\
\hline
AuFe 		& L1$_0$	   &  fcc        & 166.1  		& 165.1  		& 166.6	& 166.1	& 166.1	& 164.5	&  166.6	& 166.1		& Experiment\\
			& W2	   &  fcc	 	& 127.2		& 135.7 		& 155.6	& 161.1	& 158.2	& 158.7	&  190.2	& 185.5		& CE-SF \\
			& (001) Au$_4$Fe$_4$ SL
						& fcc	& unstable
											& 125.4		&  76.2	& 119.3		& 127.8	& 115.0		&  298.6	& 300.4     & CLDM\\
			& L1$_0^{*}$ & fcc	& 154.0		& 160.3		& 155.3	& 154.4	& 154.4		& 153.7  &  155.3	& 154.4		& CE ($\Delta H_{chem}$)\\
			& CH (``40'') 	 & fcc	&  173.5		& 155.6		& 176.3	& 174.5	& 174.5	& 174.7	&  176.3	& 174.5		&  SCE (Sec.\ \ref{sec:magnetism}) \\
\hline
AuFe$_3$ 	&L1$_2$     &  fcc		& 134.5		& 144.2		& 138.7	& 138.7	& 138.7	& 138.8	&  138.7	& 138.7		& Experiment\\
			&L1$^*_2$  &  fcc		& 134.0		& 141.9		& 137.6	& 138.2	& 139.9	& 137.4	&  139.5	& 139.9		& CE-SF,CLDM\\
\hline
\hline
\end{tabular}
\label{table:GSdH}
\end{table*}

Experimental data indicate \lot ordering at the AuFe$_3$ composition.\cite{Shield2014}  The \lot structure has the lowest energy among all structures with up to 6 atoms per unit cell that are confirmed by DFT to be fcc-stable. There are more complicated fcc-stable structures slightly lower in energy, confirmed by DFT, which can be viewed as defective L1$_2$ with a periodic arrangement of antiphase boundaries. One such structure is L1$^*_2$ with 8 atoms per unit cell, which is 0.5 meV/atom below L1$_2$. The fcc CE-SF also predicts lower-energy structures that are fcc-unstable. This problem underscores the basic limitation of the CE-SF method.
However, if we assume that the fcc-bcc transformation is blocked, there is no direct disagreement with experiment, except that the calculations suggest the possibility of the proliferation of antiphase boundaries in the L1$_2$ phase.

On the other hand, in AuFe and Au$_3$Fe the energetics of fully ordered structures is in conflict with experimental observations of \loz and \lot structures:\cite{Shield,Shield2014} We found a number of {\em fcc-stable} structures that have lower energies compared to \loz and \lott, respectively. For AuFe, the (311) Au$_2$Fe$_2$ SL (known as the W2 structure) has the lowest formation enthalpy among the DFT-confirmed fcc-stable structures, 38 meV/atom lower than \lozz. Other structures, including the (001) and (111) Au$_2$Fe$_2$ SLs (known as Z2 and V2) are also fcc-stable and have $\Delta H$ below that of \lozz. While the CE-SF predicts a large number of structures with even lower energies, those directly checked in DFT turned out to be fcc-unstable. At the Au$_3$Fe composition the lowest DFT-confirmed fcc-stable structure with up to 6 atoms/cell is the (001) Au$_3$Fe SL (known as Z1), which is more than 100 meV/atom lower than \lott. We also found several DFT-confirmed structures with 8 atoms/cell below Z1. These structures are Au$_6$Fe$_2$ superlattices with various stacking directions, such as [011], [131], [001]. All of them relax strongly towards bcc geometry.

Summarizing the results of this section, among the three experimentally suggested \loz (AuFe) and \lot (Au$_3$Fe and AuFe$_3$) structures, only \lot ordering in AuFe$_3$ is not in direct conflict with GGA energetics of fully ordered fcc-stable structures at $T=0$. Neither \lot Au$_3$Fe nor \loz AuFe has the lowest energy among the fcc-stable structures at the respective compositions, or is even close to being the lowest.

\subsection{Effects of restricting geometric relaxation}
\label{sec:ordered:CLDM}

It is instructive to compare the contributions to the formation enthalpies of the most stable fully ordered structures from the chemical interaction, local relaxations, and uniform strain. In particular, since the CLDM method neglects the uniform strain contribution, as it is justified in Appendix \ref{app:proof} for nearly disordered alloys treated in later sections, one may ask how accurate its predictions would be for the lowest-energy fully ordered structures. It is also useful to compare the numerical and systematic errors of different methods.

Table \ref{table:GSdH} lists several structures that have been fully relaxed, relaxed with a fixed unit cell (i.e., only allowing internal relaxations), or not relaxed at all (i.e., with atoms fixed at the ideal fcc positions). The formation enthalpies calculated from DFT are compared to the predictions of the different models. Table \ref{table:GSdH} includes both structures suggested by experiments and candidate structures predicted as ground states by CE-SF, S-CLDM, or chem-CE using direct enumeration of orderings with relatively small unit cells.

It is clear that structural relaxations change the energetic hierarchy. For example, the lowest-energy \emph{unrelaxed} structure at the AuFe composition, among those with up to 8 atoms per unit cell, is the tetragonal structure labeled as L1$_0^{*}$ in Fig.\ \ref{fig:structures}. In contrast, CLDM (combined with chem-CE, as discussed above) predicts that the lowest-energy fixed-cell structures at AuFe composition either are or resemble long-period (001) SLs.

The most stable structure predicted by CLDM, among about 10000 enumerated, is the A$_4$B$_4$ (001) SL, which consists of alternating 4-monolayer-thick slabs of Fe and Au stacked along the [001] direction. (This structure has one of the highest energies without the relaxation.) A DFT calculation with a fixed unit cell gives an even lower formation enthalpy for this structure, by as much as 43 meV/atom. This underestimation of the relaxation energy in CLDM is due to the collapse of the thick Fe regions in this SL towards bcc geometry. Under the fixed-cell constraint this collapse is incomplete, and the structure as a whole formally passes our fcc filter, but the fully-relaxed structure is filtered out as having relaxed away from fcc. (Even then it is not quite bcc-like: the interlayer spacings suggest nearly perfect fcc lattice within the Au$_4$ layers and distorted bcc in AuFe$_4$Au.)

The shorter-period (001) SLs, such as Au$_2$Fe$_2$ (Z2) or Au$_{1}$Fe$_{1}$ (L1$_0$), do not undergo such a drastic collapse of the Fe regions, and their DFT formation enthalpies are similar to CLDM predictions.
Similar to the fcc CE-SF, the CLDM predicts that the lowest-energy structure at the AuFe$_3$ composition is the L1$^*_2$ structure.

CLDM makes two assumptions: zero uniform strain and the harmonic approximation. The former is justified for nearly random alloys (see Appendix \ref{app:proof}), as well as for ordered structures with cubic symmetry, like \lott. Otherwise the neglect of uniform strain is an approximation. Comparing the ``fully relaxed'' and ``fixed cell'' GGA columns in Table\ \ref{table:GSdH}, we see that the zero-strain approximation works well for low-period tetragonal Z1, \lozz, L1$_0^{*}$, and CH, but fails for the orthorhombic W2 structure. The harmonic approximation works well in all cases with the exception of the (001) Au$_4$Fe$_4$ SL, which was explained above.

\section{Ordering from disordered alloys}
\label{sec:disordered}

We turn to the physics of the early ordering stages. We found in Sec.\ \ref{sec:ordered:CE} that GGA energetics of fully ordered structures conflicts with the experimental observation of L1$_0$ ordering in AuFe nanoparticles. However, if the ordering observed in slowly cooled nanoparticles is incomplete, it may rather reflect the ordering tendencies in the random alloy. Since ordering does not involve long-range diffusion, we assume that the key mechanisms are still thermodynamic, rather than kinetic, in nature.

The preferred ordering vectors in the random alloy can be different from those characterizing the fully ordered structures observed at low temperatures, as has been shown, for example, for Ni-V and Pd-V.\cite{Wolverton-SROLRO}

\subsection{Energetics of nearly-random alloys}
\label{sec:disordered:main}

As shown in Appendix \ref{app:proof}, if the alloy deviates only slightly from the disordered state, the neglect of the uniform strain in CLDM is justified to the leading order in the order parameter. The \loz and \lot ordering phase transitions are allowed by the Lifshitz criteria to be second-order, although they are usually weakly first-order. For the nanoparticles, we have assumed that the phase separation is blocked, which means the ordering develops continuously. Thus, to understand the driving forces for ordering in the core of a Au-Fe nanoparticle, we use CLDM to evaluate the ordering tendencies in a random fcc Au-Fe alloy, as a function of volume and concentration. Note that the CE-SF method is poorly suited to study the energetics in nearly-disordered Au-Fe alloys due its relatively large prediction error (see Table\ \ref{table:CEparams}) and the unreliable predictions of fcc-unstable structures.

The black lines in Fig.\ \ref{fig:vtotal} show the effective potential $J_\mathrm{eff}(\mathbf{k})$, calculated at the equilibrium volume for several concentrations, using either the real-space CE fit of the CLDM (which is accurate away from the $\Gamma$-point) or the S-CLDM with the residual CE (which captures the elastic singularity at $\Gamma$). The two curves are almost identical far from the $\Gamma$ point, while we expect S-CLDM to work better in its vicinity. The dotted red line shows the contribution from the residual CE, which is relatively small, as expected from the discussion in Appendix \ref{s-cldm} (see Fig.\ \ref{fig:refit}).

\begin{figure}[ht!]
\begin{center}
\includegraphics[width=0.45\textwidth]{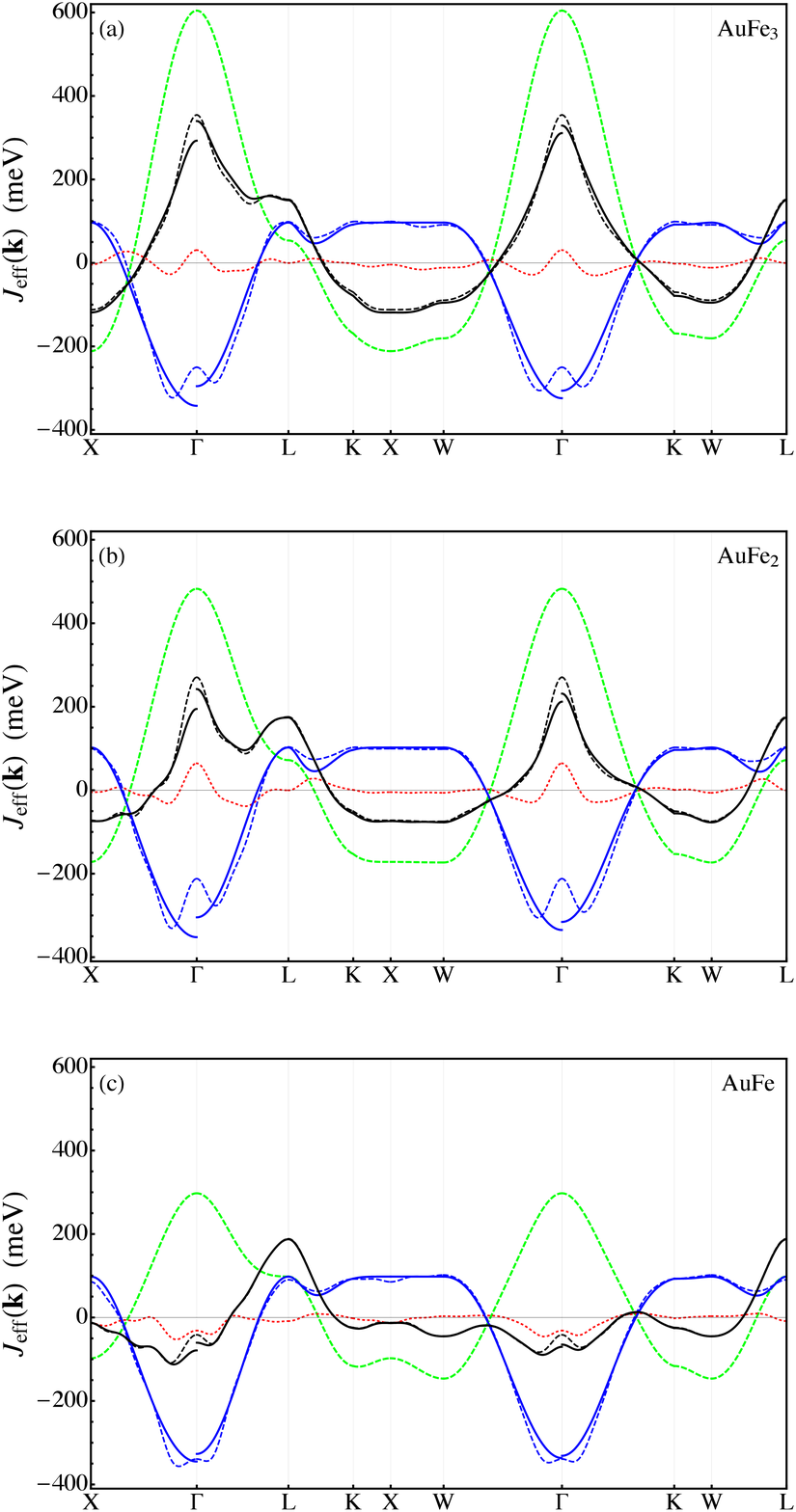}
\end{center}
\caption{
(Color online) Total  effective potential $J_\mathrm{eff}(\mathbf{k})$ (black) and the individual chemical (green) and relaxation-induced (blue) contributions, plotted along the special directions in the Brillouin zone, at the following concentrations of Au$_{1-x}$Fe$_x$:
(a) $x=0.75$, (b) $x=0.666$, (c) $x=0.5$.
Thick green line: $J_{chem}(\mathbf{k})$.
Dashed blue line: $J_\mathrm{SI}(\mathbf{k})$ from the CLDM-based CE (Appendix B).
Solid blue line: $J_\mathrm{SI}(\mathbf{k})$ from S-CLDM (first term in Eq.\ (C1)).
Dotted red line: $J_\mathrm{res}(\mathbf{k})$ from Eq.\ (C1).
Dashed black and solid thick black lines: total $J_\mathrm{eff}(\mathbf{k})$ with $J_\mathrm{SI}(\mathbf{k})$ from the CLDM-based CE and from S-CLDM, respectively.}
\label{fig:vtotal}
\end{figure}

In AuFe$_{3}$, the reported \lot ordering is characterized by ordering vectors at the three inequivalent X-points. Indeed, the black line in Fig.\ \ref{fig:vtotal}(a) shows that $J_\mathrm{eff}(\mathbf{k})$ is minimal and almost flat on the face of the Brillouin zone containing the X point, revealing ordering tendencies at the corresponding wave vectors. While this result does not conclusively point to \lot ordering, it is possible that the X point can be preferred either kinetically or due to higher-order interaction effects.

The \loz ordering reported for the equiatomic AuFe alloy is characterized by one X-point ordering vector. However, Fig.\ \ref{fig:vtotal}(c) indicates that $J_\mathrm{eff}(\mathrm{X})$ is close to zero at this composition, indicating the {\em absence} of a thermodynamic driving force for X-point ordering in AuFe. The global minimum of the total $J_\mathrm{eff}(\mathbf{k})$ is reached close to the zone center on the $\Gamma$X line. There is also a secondary minimum approximately half-way between $\Gamma$ and X. These features are consistent with our findings for the fully ordered structures discussed in Sec.\ \ref{sec:ordered:CE}.

We conclude that the calculated ordering tendencies are consistent with partial \lot ordering in AuFe$_3$  \cite{Shield2014} but not with \loz ordering in AuFe.\cite{Shield,Shield2014} A similar analysis could not be performed for Au$_{3}$Fe, because the CLDM predicts the random alloy to be dynamically unstable at this composition, indicating the presence of large anharmonic distortions.

\subsection{Effects of surface segregation and surface tension on ordering in nanoparticles}
\label{sec:disordered:volume}

The blue and green lines in Fig.\ \ref{fig:vtotal} show the  $J_{chem}(\mathbf{k})$ and $J_\mathrm{SI}(\mathbf{k})$ contributions to the effective potential $J_\mathrm{eff}(\mathbf{k})$ [see Eq.\ (\ref{eq:Jchem})]. There is a strong competition between ordering and phase separation tendencies introduced by the ``chemical'' and strain-induced interactions; they have opposite signs throughout most of the Brillouin zone. Due to this competition, the ordering tendencies may be sensitive to pressure and alloy composition.

Among all the transition metals, Au is one of the strongest surfactants when alloyed with fcc Fe.\cite{Ruban-seg} A strong enrichment of the nanoparticle surface by Au, and of its core by Fe, may therefore be expected. Comparison of the three panels of Fig.\ \ref{fig:vtotal} shows that enrichment by Fe beyond the equiatomic composition gradually stabilizes the X-point ordering. The minimum along the $\Gamma-$X line shifts from the vicinity of $\Gamma$ in AuFe to an almost flat section near X in AuFe$_{2}$, and further to a clear minimum at X in AuFe$_{3}$. A similar trend is seen along other directions leading to the X-point. Thus, X-point ordering may become preferred at $x\gtrsim2/3$, similar to the $x=0.75$ case considered above. The enrichment of the nanoparticle core by Fe could, therefore, explain the observation of L1$_0$-type ordering in AuFe nanoparticles. \cite{Shield}

Such enrichment is also suggested by unexpectedly small measured lattice parameters of \lot and \lozz. Table \ref{table:alat} compares them with calculations for fully ordered \lot and \loz and random alloys; the latter are estimated from a separate CE of the atomic volume of fcc-based structures (V-CE). The experimental lattice constants are systematically smaller than the GGA values, by as much as 5-8\% in AuFe, which is much more than is typical for GGA.\cite{HaasBlaha2009} The experimental value for \loz AuFe is also smaller than expected from the comparison with similar alloys. For example, L1$_0$-ordered FePt has $a=3.85$ \AA, and the lattice parameter of Au is 0.15 \AA\ larger than Pt; a Vegard-law estimate then gives 3.9--3.95 \AA\ for AuFe even before accounting for the positive deviation from the Vegard law, which is expected for phase-separating alloys and confirmed by calculations.

The observed reduction of the lattice parameter is not fully explained by Fe enrichment, because this would require the nanoparticle core to contain more than 75\% Fe. At this composition the minority Fe atoms in the Au layer of the \loz structure reduce the L1$_0$ order parameter to less than 0.5, which is at odds with the excellent matching between the experimental $c/a$ ratio and the DFT result for the \emph{fully ordered} \loz (Table \ref{table:alat}). Surface tension is another possible source of lattice contraction. Given the typical surface energy of order $1$ eV/atom, the excess pressure due to the surface curvature is only a few kilobars for a 10 nm radius, which should reduce the lattice constant by much less than a percent.

Additional compressive surface stress, unrelated to curvature, may develop in metals at the end of the $d$-series \cite{Takeuchi1989} due to the spill-out of the $s$ electrons into the vacuum and the resulting stronger bonding of the $d$ electrons. In gold this stress is particularly strong due to relativistic effects, \cite{Takeuchi1989} which was argued to be the cause of the surface reconstruction. It may be energetically more favorable to relieve this stress by contracting the nanoparticle core, rather than by reconstructing the surface. While we did not attempt to further quantify the contributions of the surface segregation and surface-induced stress mechanisms, the substantial observed contraction of the nanoparticle core suggests that both Fe enrichment and surface-induced stress may be contributing.

\begin{table*}[htb]
\caption{Theoretical and experimental lattice parameters of \lot and \loz structures and random alloys. Calculations are from GGA, with V-CE for random alloys.}
\begin{tabular}{ll | c c c c | l}
\hline
\hline
\multirow{2}{*}{Ordering}
			& \multirow{2}{*}{Composition}
							&  \multicolumn{2}{c}{Theory}  	& \multicolumn{2}{c|}{Experiment}  & \multirow{2}{*}{Reference}\\
\cline{3-6}
 			&			 		& $a$ (\AA) &$c/a$ & $a$ (\AA) & $c/a$  \\
\hline
Au$_3$Fe \lot   &  Fe$_{0.25}$Au$_{0.75}$  	&	4.083    	&	 1       		&			&	   & \\
			&Fe$_{0.33}$Au$_{0.67}$	&			&			&	3.71		&		1	   & Ref. \onlinecite{Shield2014} \\
\cline{1-2}
Random      	&  Fe$_{0.25}$Au$_{0.75}	$	&	4.071    	&  1 		&			&			   & \\
 		      	&  Fe$_{0.33}$Au$_{0.67}	$	&	4.030    	&  1 		&			&			   & \\
\hline
AuFe L1$_0$     &  Fe$_{0.5}$Au$_{0.5}	$	&	3.965    	&  0.980 		&			&		   & \\
			 & Fe$_{0.511}$Au$_{0.504}$	&		&			&	  3.67       	&   0.981     & Ref. \onlinecite{Shield} \\
			 & Fe$_{0.53}$Au$_{0.47}$	&			&			&	  3.67       	&   0.981     & Ref. \onlinecite{Shield2014} (HRTEM) \\
			 & Fe$_{0.53}$Au$_{0.47}$	&			&			&	  3.74       	&   0.962     & Ref. \onlinecite{Shield2014} (SAED) \\
\cline{1-2}
Random      	&  Fe$_{0.5}$Au$_{0.5}	$	&	3.939    	&  1 		&			&			   & \\
\hline
AuFe$_3$ \lot	& Fe$_{0.75}$Au$_{0.25}$  	&	 3.776    	&   1      		&			&		  & \\
			 &Fe$_{0.79}$Au$_{0.21}$ 	&			&			&	3.65 	 	&   1    	  &  Ref. \onlinecite{Shield2014}         \\
\cline{1-2}
Random      	&  Fe$_{0.75}$Au$_{0.25}	$	&	3.786    	&  1 		&			&			   & \\	
\hline
\hline
\end{tabular}
\label{table:alat}
\end{table*}

Fig.\ \ref{fig:volume} shows the dependence of $J_\mathrm{eff}(\mathbf{k})$ in the equiatomic alloy on the lattice parameter $a$ used to construct the CLDM, which is subsequently approximated by S-CLDM. As shown in Appendix D, calculations at fixed $a$ produce $J_\mathrm{eff}(\mathbf{k})$ that is appropriate for isobaric conditions in the core of a nanoparticle. Non-zero pressure could be due to surface tension or coherency stress from surface segregation. It is clear that within a fairly large range of the lattice parameter (3.7-3.95 \AA) the minimum of $J_\mathrm{eff}(\mathbf{k})$ is reached at the same point on the $\Gamma$X line, and there is no significant trend toward the stabilization of the X-point ordering. Thus, Fe enrichment of the nanoparticle core appears to be critical for the \loz order to develop.

\begin{figure}[hbt]
\begin{center}
\includegraphics[width=0.45\textwidth]{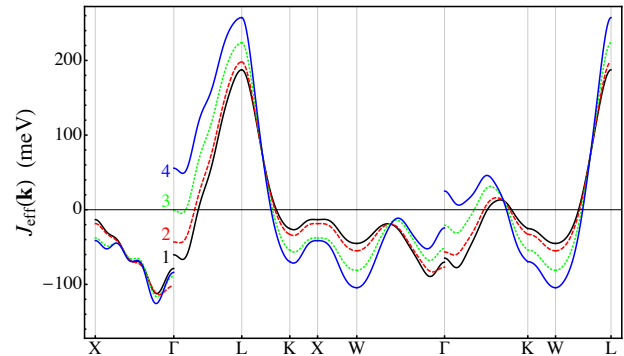}
\end{center}
\caption{(Color online) Total effective potential $J_\mathrm{eff}(\mathbf{k})$ in AuFe from S-CLDM (Eq.\ (C1)) at different volumes. Black, red, green, and blue lines (labeled 1, 2, 3, 4): $a=3.953$ \AA\ (equilibrium at zero pressure); 3.901 \AA; 3.8 \AA; 3.7 \AA.}
\label{fig:volume}
\end{figure}

To conclude this section, L1$_0$-type ordering in nominally equiatomic AuFe nanoparticles \cite{Shield} may be caused by strong Fe enrichment of the nanoparticle core, which promotes X-point ordering tendencies; indeed, such enrichment is expected theoretically and indirectly evidenced experimentally.

\subsection{Spinodal stability of fcc alloys}
\label{sec:disordered:spinodal}

As mentioned in Sec.\ \ref{sec:ordered:CE}, the (001) Au$_2$Fe$_2$ SL (known as Z2) is fcc-stable and has $\Delta H$ below that of \lozz, which is a (001) Au$_1$Fe$_1$ SL. This tendency is also seen for other SL directions in both fcc and bcc structures: Au$_2$Fe$_2$ SLs tend to have lower energy than the corresponding Au$_1$Fe$_1$ SLs. The (001) Au$_4$Fe$_4$ SL combining fcc-like and bcc-like regions has an even lower energy (Sec.\ \ref{sec:ordered:CLDM}). This suggests that the Au-Fe system may be unstable with respect to spinodal decomposition. The driving force for such decomposition is suggested by the downward curvature of the $\Delta H(x)$ line for the random alloy in Fig.\ \ref{fig:DeltaHunconstrained}, which at low temperatures approximates the second derivative of the free energy. However, since the spinodal decomposition produces a coherently strained two-phase mixture, the positive contribution of the coherency strain needs to be added to this second derivative.

We estimate the coherency strain contribution assuming the random alloy is elastically isotropic, with composition-independent Young's modulus $E$ and Poisson's ratio $\nu$. The concentration dependence of the lattice parameter of the random alloy $a(x)$ was obtained from the V-CE. Let the alloy be decomposed into planar regions of equal volumes with compositions $x_i=x_0\pm \delta x$. To the lowest order, the second derivative $F^{\prime\prime}=\partial^2F/\partial(\delta x)^2$ is
\begin{align}
    F^{\prime\prime}=F_0^{\prime\prime}+
	\frac{2EV}{1-\nu} \left(\frac{d\ln a}{dx}\right)^2 ,
\end{align}
where $F_0^{\prime\prime}$ excludes the coherency strain contribution.

We find that in a wide composition range the coherency strain nearly cancels the negative curvature of the incoherent formation enthalpy, i.e., the driving force for spinodal decomposition is eliminated already at $T=0$.
Taking $E=1.8$ Mbar and $\nu=0.27$, \cite{moduli} we find that the random alloy remains spinodally stable everywhere, with the smallest $F^{\prime\prime}$ around 70\% Fe. A smaller value $E=1.4$ Mbar, obtained by averaging the elastic moduli of the 12-atom Au$_4$Fe$_8$ SQS, \cite{SQS} leads to a wide marginally unstable region. The above is a crude estimate, which is also sensitive to CE details, and we expect S-CLDM should provide a more reliable estimate by fully capturing the harmonic part of the strain-induced contribution at the concentrations when it is available. Nevertheless, it provides an insight into the spinodal stability at the Au-rich compositions, where the random alloy is dynamically unstable and CLDM can not be used. The CE-based estimates become more robust at $x_\text{Fe}\lesssim 0.25$ and indicate the lack of any spinodal instability at these compositions.

In S-CLDM the spinodal instability corresponds to a negative value of $J_\mathrm{eff}$ at the $\Gamma$-point. From Fig.\ \ref{fig:vtotal}, we see that both the chemical and the elastic contributions to $J_\mathrm{eff}$ increase manifold as ${\bf k} \rightarrow \Gamma$ and indeed largely cancel each other. At the equiatomic AuFe composition the negative elastic contribution prevails, although the minimum $J_\mathrm{eff}({\bf k})$ value corresponds to long-period SLs rather than the fully decomposed alloy, which would be favored if $J_\mathrm{eff}({\bf k})$ had its global minimum at $\Gamma$. Increasing the Fe content in the random alloy rapidly removes the spinodal instability, and the AuFe$_{2}$ alloys are clearly stable.

In summary, Au-Fe alloys are spinodally unstable at equiatomic composition, but become stable at least for $x_\text{Fe}\lesssim 0.25$ and $x_\text{Fe}\gtrsim 2/3$. Even in the nominally equiatomic nanoparticles the strong surface segregation of Au and the enrichment of the core by Fe (discussed in Sec.\ \ref{sec:disordered:volume}) are likely to spinodally stabilize both the nanoparticle core and the Au-rich surface. Moreover, even at $x=0.5$ the spinodal instability is weaker than the ordering instability at the finite $\mathbf{k}$ vector where $J_\mathrm{eff}({\bf k})$ has its global minimum.

\section{Effects of magnetism on chemical ordering}
\label{sec:magnetism}

The results presented above were obtained under the assumption that the alloys are always ferromagnetically ordered. We now consider the validity of this assumption and the consequences of relaxing it.

Fig.\ \ref{fig:cpadh} shows the formation enthalpies of the FM and PM states calculated using CPA. It shows that the magnetic coupling in disordered fcc Au-Fe alloys is strongly ferromagnetic as long as the concentration of Au it not too small. The increase in the lattice parameter due to the Au size effect removes the magnetic frustration characteristic for pure fcc Fe. At the 50\% concentration the energy difference $E_{mag}$ between the PM and FM states is 80 meV/atom, and the mean-field Curie temperature $T_C$ of 1240 K is comparable to the FePt and FePd alloys.\cite{Staunton2006}

\begin{figure}[hbt]
\includegraphics[width=0.45\textwidth]{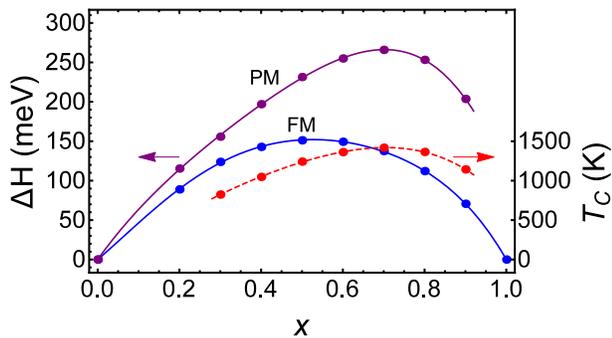}
\caption{(Color online) Solid lines: formation enthalpy of the random Au$_{1-x}$Fe$_x$ alloy in the ferromagnetic (FM) and paramagnetic (PM) states from the coherent potential approximation (CPA) calculations (with disordered local moments (DLM) for the PM state).  Dashed line:  the mean-field Curie temperature (right axis).}
\label{fig:cpadh}
\end{figure}

Measurements in Au-Fe films \cite{McGuire} have found the Curie temperature in the 550--600 K range at the 50\% concentration. However, the magnetic moment of Fe in those films ($2.2 \mu_B$) was smaller compared to $2.9 \mu_B$ found in an earlier study. \cite{Felsch} The local moment in our calculations is close to the latter value, which is also characteristic for FePt and FePd. This discrepancy in the local moment and $T_C$ suggests that the films studied in Ref.\ \onlinecite{McGuire} could have been highly defective and far from the ideal fcc structure.

To assess the effects of magnetic disorder on the ordering tendencies, we chose several ordered equiatomic structures and constructed a separate quasi-binary (Ising) spin-cluster expansion (SCE) for each of them. The choice of the SCE interaction parameters in each structure was restricted only by crystallographic and time-reversal symmetry. The nearest-neighbor exchange couplings are universally ferromagnetic and dominant in all structures, other couplings being smaller by at least a factor of 5.

Three SCE versions (indexed by label $n=0$, 1, or 2) were constructed for each structure $\sigma$, with the fixed unit cell, and with the cell-internal atomic positions: (0) kept at the ideal fcc lattice, (1) relaxed in the FM state and then used for all other spin configurations, and (2) relaxed independently for each  magnetic configuration. An estimate $E^{(n)}_\mathrm{PM}(\sigma)$ for the PM energy was obtained from each SCE by setting all spin correlators to zero. We have also considered one 16-atom SQS, \cite{SQS} but, instead of building a SCE, we simply averaged its energy over three randomly assigned magnetic configurations.

The unrelaxed magnetic energy $E_{mag}=E^{(0)}_\mathrm{PM}-E^{(0)}_\mathrm{FM}$ represents the magnetic contribution to $\Delta H_{chem}$ of the PM state. We can further define two estimates of the relaxation energy in the PM state, $E^{(n)}_{rel}=E^{(n)}_\mathrm{PM}-E^{(0)}_\mathrm{PM}$ ($n=1,2$), which can be compared with the FM state. The results are listed in Table \ref{table:magn}, and we can now estimate the effect of magnetic disorder on $J_\mathrm{SI}(\mathbf{k})$ and $J_{chem}(\mathbf{k})$.

\begin{table}[htb]
\caption{Magnetic energy $E_{mag}$ and the relaxation energies for the FM and PM states (the latter evaluated in two ways: see text). Energies are in meV/atom; $a=3.9$ \AA.}
\begin{tabular}{| l l |c|c|c|c|}
\hline
\multirow{2}{*}{Structure}	&Superlattice    & \multirow{2}{*}{$E_{mag}$} & \multicolumn{3}{c|}{$E_{rel}$}\\
\cline{4-6}
                   			&  direction     &		& FM        &   PM (1)  &   PM (2)     \\
\hline
L1$_0$                  & AB (001) 			& 84.8      &   0       &   0       &   13.0    \\
L1$_1$                  & AB (111) 			& 77.4      &   0       &   0       &   15.3    \\
Y2 			& A$_2$B$_2$ (011)     & 83.2      &   45.3    &   53.3    &   61.4    \\
CH 			& A$_2$B$_2$ (012)     & 46.3      &   0       &   0       &   10.1    \\
W2 			& A$_2$B$_2$ (113)     & 73.3      &   36.5    &   34.5    &   44.7    \\
``101'' 	& A$_2$B$_2$AB (135)  & 61.5      &   11.4    &   11.6    &   21.8    \\
SQS-16   	&  			              & 86.6      &   62.2    &   60.2    &   75.6    \\
\hline
\end{tabular}
\label{table:magn}
\end{table}

First, we observe that $E_{rel}$ is similar in the FM and PM states for all structures listed in Table \ref{table:magn}. Allowing the lattice to relax for each magnetic configuration increases the relaxation energy by 8-16 meV/atom for all structures, including those that do not relax at all in the FM configuration. Thus, magnetic disorder has little effect on $J_\mathrm{SI}(\mathbf{k})$.

$E_{mag}$ for the unrelaxed SQS-16 (87 meV/atom) is similar to the CPA value (80 meV/atom), confirming our estimate of the Curie temperature in the random alloy. $E_{mag}$ in the L1$_0$ structure is almost identical to the SQS, and it is also quite similar for most other structures. A notable exception is the W-point-ordered (012) A$_2$B$_2$ SL (known as the CH structure, or structure ``40'' in the notation of Ref.\ \onlinecite{Kanamori}), where $E_{mag}$ is reduced to 46 meV/atom. This suggests that in the PM state $J_{chem}(\mathbf{k})$, and thus $J_\mathrm{eff}(\mathbf{k})$, should be significantly lowered near the W point. A preference for W-point ordering was also found in an earlier CPA-based study \cite{Ling} for PM Au-rich Au-Fe alloys. (CPA describes $J_{chem}$ while disregarding $J_\mathrm{SI}$.)

The CH structure is illustrated in Fig.\ \ref{fig:structures}. It is a superposition of two W-point concentration waves:
\begin{equation}
\sigma_i = \frac12(1+i)\exp(i\mathbf{Q}_1 \mathbf{R}_i)+\frac12(1-i)\exp(i\mathbf{Q}_2 \mathbf{R}_i),
\end{equation}
where one can take $\mathbf{Q}_1=2\pi(1/2,1,0)/a$ and $\mathbf{Q}_2=2\pi(1/2,0,1)/a$. Similar to \loz and \lott, it is one of only nine fcc structures \cite{Chepulskii2012} that, according to the Lifshitz criteria, may order through a second-order transition. Its energy in the fully relaxed FM state (Table\ \ref{table:GSdH}) is 10 meV/atom above \loz and 50 meV/atom above W2, which is the lowest-energy identified AuFe fcc structure. (Confusingly, W2 has no relation to the W point).

For the PM state, approximating the interaction as purely pairwise, we estimate that a 40 meV/atom reduction in the energy of CH translates into a 80 meV reduction of $J_{chem}(\mathrm{W})$. Thus, at high temperatures W-point ordering could compete with the ordering tendencies identified in the previous sections, and could potentially be observed if the samples were quickly quenched.

In Sec.\ \ref{sec:magnetismBain}, we demonstrated that two specific AFM orderings with antiparallel nearest neighbors render the \loz phase dynamically unstable in the fcc geometry. An exhaustive enumeration of all periodic structures with up to 16 atoms per unit cell with the SCE version (0) showed that FM state is the magnetic ground state for \lozz. Even allowing cell-shape relaxation, i.e., letting the fcc-unstable structures reach their optimal geometry, does not reveal possible competing spin orders. We have also calculated the energy of AFM \lot Au$_3$Fe (with all nearest Fe spins anti-aligned, corresponding to rocksalt-type spin ordering), and found that its energy is 1.9 meV/atom (7.6 meV/Fe) {\em higher} compared to the FM ordering. These results are in contrast with earlier calculations \cite{Shield2014} suggesting that an AFM state is closely competitive in \loz AuFe and favorable in \lot Au$_3$Fe.

\section{Discussion and Conclusions}
\label{sec:Discussion}

It is not unusual for ordering tendencies in nearly-random alloys to differ from ground-state ordering at the same composition, as in Ni-V and Pd-V systems. \cite{Wolverton-SROLRO}
However, our finding that the experimentally observed ordering strongly contradicts the DFT results for fully ordered structures (cf. Table\ \ref{table:GSdH}) but is consistent with weak ordering tendencies (provided additional Fe-enrichment of the
nanoparticle core, in the case of AuFe) is still surprising.
First, in all the previously reported cases where the
ordering tendencies in the fully- and ``weakly''-ordered alloys were inconsistent, the ``weak'' ordering was the case of a short-range order. In the case of Au-Fe, the experimental procedures employed in observing the ordering in Au-Fe nanoparticles suggest it be classified as long-range (even though formally, there is no distinction between short and long-range order in finite-size nanoparticles).
Second, the quantitative size of the energy penalty of forming fully ordered \lot and \loz compared to alternative orderings (Table\ \ref{table:GSdH}) is  surprising. It has recently been demonstrated\cite{beta2hybrid}
that the semi-local density functionals, such as GGA-PBE, do not always capture correctly the energetic hierarchy of metal alloy structures;
in particular, some conflicts between the DFT predictions and experiment known for Cu-Au alloys\cite{Ozolins98} disappear
if the more accurate, non-local hybrid HSE functional is used.\cite{beta2hybrid})
The largest
(among nine structures in three alloy systems) hybrid correction to PBE $\Delta H$ has been reported for CuAu \loz and constitutes 35meV/atom.\cite{beta2hybrid}
It is possible that in Au-Fe, the energy of \loz and \lot structures would be lowered by similar corrections. The ordering energetics of the alloys containing $5d$ metals may also be affected by the spin-orbit coupling.\cite{SluiterAuPd,BarabashAuPd,note:AuPd}
Due to a need of high k-mesh sampling and the slow convergence of our PBE calculations in the magnetic Au-Fe alloys, we did not attempt repeating them with the computationally much more demanding hybrid functional or spin-orbit methods.
However, both the hybrid and the spin-orbit corrections to $\Delta H$ values of different structures of the same alloy are usually of the same sign and, for similar compositions, of somewhat comparable magnitude, and thus the structural hierarchy {\em at a particular composition} is affected to a lesser degree.
Thus, we expect that the key qualitative findings of this study would not be affected by non-local and spin-orbit corrections.

In conclusion, by using {\em ab initio}-based effective Hamiltonian techniques combined with direct density functional theory (DFT) calculations, we find that the phase separation tendency in Au-Fe is inherent (rather than caused by the freedom to form distinct fcc- and bcc-based phases), and that the temperature- and composition-dependent trends such as the formation of the bcc phase in low-temperature deposition and the transformation into the fcc phase with annealing are consistent with DFT. In Fe-rich nanoparticles, the absence of the inverse transformation might be an additional evidence of ordering during anneal. On the other hand, our DFT results contradict the assumption of developing fully ordered L1$_0$ AuFe, L1$_2$ Au$_3$Fe or L1$_2$ AuFe$_3$ phases: we find more stable (lower energy) structures at each of these three compositions.
By analyzing the ordering tendencies in a nearly-random alloy, we find that they are consistent with incipient L1$_2$ order at AuFe$_3$ composition, thus being substantially different from those in fully ordered alloys. However, AuFe does not exhibit L1$_0$-type ordering tendencies even if assumed nearly-random, and instead is prone to a spinodal decomposition. We argue that the experimental lattice constant of nominally AuFe nanoparticles is too low, evidencing a substantial enrichment of nanoparticle core by Fe. Once the Fe enrichment is taken into account, the ordering tendencies become consistent with incipient L1$_0$ order and the spinodal instability is removed.
Finally, we demonstrate that the effect of surface tension on ordering tendencies is negligible, while changing the magnetic ordering may affect both the structural hierarchy and the fcc-bcc transformation pathway. The magnetic ordering is expected to occur at a fairly high temperature, and the ground states of both \loz AuFe and \lot Au$_3$Fe are predicted to be ferromagnetic.

\begin{acknowledgments}

This work was supported by the DOE EPSCoR State and National Laboratory Partnership Program under Grant No.\ DE-SC0001269. Computations were performed utilizing the Holland Computing Center of the University of Nebraska.

\end{acknowledgments}

\appendix
\section{CE details}
\label{appendix:CE}

We construct bulk fcc and bcc CEs using the ATAT package,\cite{ATAT} separately for fcc- and bcc-based structures.  The number and type of ECIs [i.e. the terms to be kept in the expansion given by Eq.\ (\ref{eq:CE})] is chosen so as to minimize the error in the {\em predicted} \mdH for structures that were not used for fitting ECI values. The latter error is estimated by the leave-one-out cross-validation (CV) error as calculated by ATAT.\cite{ATAT}

The CE approach is often modified by treating the strain-induced term separately. We use this idea in Sec.\ \ref{CLDMmethod} for the CLDM construction, but do not employ this approach in the CE-SF study (in part, due to the dynamic instability of fcc Fe at $T=0$, leading to some practical issues). We do, however, account for the coherency strain contribution, in Sec.\ \ref{sec:disordered:spinodal}, when we discuss the implications of the CE-SF results for the spinodal stability of coherent alloys.

For structural filtering, we start with the procedure of Refs.\ \onlinecite{BarabashFeX,Chepulskii2012}.
Specifically, for each structure $\sigma$, a degree $s^{(\alpha)}(\sigma)$ of its proximity to the underlying lattice type $\alpha$
($\alpha$= fcc or bcc) is defined as
\beq
s^{(\alpha)} (\sigma)^{-1}=
\frac 1 N_0
\sum\limits_{i<N_0} \sum\limits_{j\ne i}
 [ d_{ij} (\sigma)-d_{ij} (\alpha)]^2
e^{-\eta d_{ij} (\alpha)},
\label{eq:NNScore}
\eeq
where the first sum is over the $N_0$ atoms in the unit cell of the periodic structure $\sigma$ while
the second sum is over all atoms $j$ in the lattice (in practice limited to a finite portion
of the lattice by the cutoff $\eta \sim 1$).
The correspondence between the dimensionless (volume-rescaled) interatomic distances $d_{ij} $
 in the relaxed structure $\sigma$ and those
in a perfect lattice $\alpha $ is defined  by sorting the distances $\{d_{ij}(\sigma )\}$ in increasing order.
If, for example, $s^{\text{fcc}} (\sigma ) \gg s^{\text{bcc}}(\sigma )$  for a given relaxed structure $\sigma$, one may conclude that the structure has relaxed to an fcc-like geometry.
Such $s^{\text{fcc}} $ \emph{vs} $s^{\text{bcc}}$ comparison is visualized by plotting
\beq
r(\sigma)= \frac 2 \pi \arctan\left[\log\left(s^{\text{fcc}} (\sigma ) / s^{\text{bcc}}(\sigma ) \right)\right],
\label{eq:NNScoreR}
\eeq
which changes from 1 for ideal fcc to $-1$ for ideal bcc structure.

We did not employ the HS/LS filtering used in Refs.\ \onlinecite{BarabashFeX,Chepulskii2012}, because we expect the Au-Fe alloys to remain in the HS state, as discussed in Sec.\ \ref{sec:DFT}. For the CE-SF studies, we restrict consideration to HS structures with FM spin order, and analyze the effects of spin ordering separately in Sec.\ \ref{sec:magnetismBain} and \ref{sec:magnetism}.

As initial inputs, we took all possible structures up to 6 atoms per unit cell for both lattices. Those that transform to a different lattice type after the relaxation, i.e., for which $r(\sigma)$ changes sign, were excluded from the input sets. In addition, as discussed in Section \ref{sec:fccbcc:ordered}, two or more initial structures of the same lattice type sometimes relax to the same ``unmappable'' final structure, as confirmed by their final total energy, volume, and both fcc and bcc scores being equal up to a small tolerance (0.2 meV/atom, 0.1\%, and 0.01, respectively). We have examined several of the lowest-energy structures among them and found that they are hybrid bcc/fcc SLs with alternating layers of bcc-like Fe and fcc-like Au. These structures can not be unambiguously assigned to a specific configuration of the Ising model. Therefore, all unmappable structures have been excluded from the input sets. This exclusion results in a considerable reduction of the CV scores, especially in the bcc case.

Table \ref{table:CEparams} lists the total numbers of structures that were calculated in DFT, excluded, and retained as inputs for the respective bcc and fcc CE-SFs. The table also includes the number of effective cluster interactions (ECIs) and the resulting accuracy of the CE-SF predictions assessed by the CV score.

\begin{table}[htp]
\begin{center}
\begin{tabular}{c | c c c | c c c | c}
\hline
Lattice    & \multicolumn{3}{c|}{DFT input structures} & \multicolumn{3}{c|}{Number of ECIs} & CV\\
& Total & Excluded  & Used &   Pairs  & 3-body  & 4-body &  (meV) \\
\hline
fcc           & 137        & 51 & 86         &        6  & 1 &  0 & 18.8\\
bcc	       & 137        & 79 & 58         &        13 &  1  & 0 & 13  \\
\end{tabular}
\end{center}
\caption{Parameters of the cluster expansions with structural filters based on the energies of fully relaxed structures.}
\label{table:CEparams}
\end{table}

\section{Construction of the CLDM}
\label{app:CLDM}

The ``chemical'' $\Delta H_{chem}$ values are calculated for a set of 311 structures at each lattice parameter. This set has been chosen to systematically contain structures at each of the following concentrations: 25, 50, 66.7 and 75\% Fe. Specifically, it includes all structures with up to 8 atoms per unit cell for 25\%, 50\% and 75\% Fe and up to 9 atoms per unit cell for 66.7\% Fe. The calculated energies have been used to construct real-space cluster expansions for each of the lattice parameters by means of the ATAT code.\cite{ATAT} We used the same number and type of clusters for each expansion to reduce systematic errors. The basis sets and overall quality of these expansions are displayed in Table \ref{tab:clexp}.

\begin{table}[hbt]
\caption{Parameters of the CE used in the calculations. The fourth column lists the number of 2-body, 3-body and 4-body ECIs in the basis set. The CV and misfit are given in meV/atom. Rows marked C: CE for the $\Delta H_{chem}$; R: CE refits of $\Delta H_{rel}$ from CLDM; Res: CE fits for the residual errors of S-CLDM.}
\begin{tabular}{|l|c|c|c|c|c|}
\hline
\multicolumn{1}{|c|}{Alloy} & Term     & Inputs & ECIs & CV & Misfit\\

\hline
\multirow{3}{*}{Au$_3$Fe} & C   & 309	   & 5,20,85         &3.3    &1.5   \\
\cline{2-6}
			                   & R   & \multirow{2}{*} { 323}  & \multirow{2}{*} { 21,23,35 }    &3.6     & 2.8    \\
			                    &Res   &    &   	           & 3.8     &2.9   \\

\hline
\multirow{3}{*}{AuFe 3.953} & C   & 311	   & 5,20,85		           &4.0    &1.8  \\
\cline{2-6}
			                   & R   & \multirow{2}{*} {1914}  & \multirow{2}{*} {39,50,35}    &1.8    & 1.7  \\
			                    &Res   &    & 	           & 1.7   &1.6  \\
\hline
\multirow{3}{*}{AuFe 3.901}     & C   & 311	   & 5,20,85		           &4.3    &1.8  \\
\cline{2-6}
			                   & R   & \multirow{2}{*} {1936}  & \multirow{2}{*} {39,50,35}    &1.9    &1.8  \\
			                    &Res   &    & 	           & 1.6   &1.5  \\
\hline
\multirow{3}{*}{AuFe 3.8}      & C   & 311	   & 5,20,85		           &6.1    &2.0  \\
\cline{2-6}
			                   & R   & \multirow{2}{*} {1936}  & \multirow{2}{*} {39,50,35}    & 2.3   &2.1  \\
			                    &Res   &    & 	           & 1.8   &1.7  \\
\hline
\multirow{3}{*}{AuFe 3.7}       & C   & 311	   & 5,20,85		           &7.0    & 2.0 \\
\cline{2-6}
			                   & R   & \multirow{2}{*} {1936}  & \multirow{2}{*} {39,50,35}    &3.0    & 2.8 \\
			                    &Res   &    & 	           &  2.2  &2.0  \\
\hline
\multirow{3}{*}{AuFe$_2$}            & C   & 311	   & 5,20,85		           &4.6    &1.9  \\
\cline{2-6}
			                   & R   & \multirow{2}{*} {980}  & \multirow{2}{*} {39,50,35}    & 1.5   &1.3  \\
			                    &Res   &    & 	           &1.2    &1.0  \\
\hline
\multirow{3}{*}{AuFe$_3$}            & C   & 311	   & 5,20,85		           &6.0    &2.0  \\
\cline{2-6}
			                   & R   & \multirow{2}{*} {440}  & \multirow{2}{*} {39,50,35}    &1.7    &0.9  \\
			                    &Res   &    & 	           &1.2    &0.7  \\
\hline
\end{tabular}
\label{tab:clexp}
\end{table}

For the Kanzaki forces $\mathbf{F}_i$ we use the expansion:\cite{CLDM}
\begin{align}\label{fexp}
\mathbf{F}_i&=\sum_{j=nn(i)} \left(f_1+\bar f_1\sigma_i\right)\sigma_j\,\mathbf{e}_{ji}
\nonumber\\
&+\sum_{jk}[(f_t+\sigma_i\bar f_t)\eta_{ijk}+(f_l+\sigma_i\bar f_l)\zeta_{ijk}]\sigma_j\sigma_k\mathbf{e}^i_{jk},
\end{align}
where $\sigma_i=1$ for Au and $-1$ for Fe. In the first sum $\mathbf{e}_{ji}$ is the unit vector pointing from site $j$ toward site $i$. The second sum corresponds to coplanar forces from two-site clusters $P=\{j,k\}$, where $\mathbf{e}^i_{jk}$ is a unit vector pointing from the midpoint between $j$ and $k$ towards $i$, and $\eta_{ijk}$ and $\zeta_{ijk}$ are projectors selecting specific cluster shapes. Namely, $\eta_{ijk}=1$ only if $i$, $j$ and $k$ make a triangle of nearest neighbors; $\zeta_{ijk}=1$ only if $i$, $j$, $k$ form a two-link straight chain of nearest neighbors with $i$ at an end.

Table \ref{forcepars} shows two fitted sets of the Kanzaki force parameters, separated by a horizontal line. The first set contains only the two nearest-neighbor terms $f_1$ and $\bar f_1$, and the second set includes all parameters appearing in Eq.\ (\ref{fexp}). Similar to the Cu-Au and Fe-Pt systems considered previously,\cite{CLDM} the simpler two-parameter expansion already provides a reasonably good fit, while the additional parameters further improve its quality. The parameters $f_1$ and $\bar f_1$ remain dominant in the extended fit.

\begin{table}[htb]
\caption{Parameters of the cluster expansion for the Kanzaki forces (meV/\AA). The terms are defined in Ref.\ \onlinecite{CLDM}. The horizontal line separates two different fittings.}
\begin{tabular}{|c|c|c|c|c|c|c|}
\hline
$x_\mathrm{Au}$     &   1/2   &  1/2   &   1/2  &   1/2  &   1/3    &   1/4 \\
$a$, \AA            &   3.953 & 3.901  &   3.8  &   3.7  &  3.862   &  3.811    \\
\hline
$f_1$            & 252     &   295  &   398  &   536  &  310     &   333     \\
$\bar f_1$     & 85     &   100  &   135  &   182  &  108     &   119     \\
\hline
$f_1$            & 255    &   298  &   401  &   538  &  309     &   351     \\
$\bar f_1$     & 80      &    94  &   129  &   174  &  128     &   171     \\
$f_2$            & 12.3    &   13.9 &    18  &    24  &  21.7    &   26.1    \\
$\bar f_2$     & $-6.1$  &$-6.8$  &$-9$    &$-10$   & $-9.3$   &   $-9.0$  \\
$f_3$            & 0.2     & 0.9    & 2.5    &  4.5   & $-2.3$   &   $-3.8$  \\
$\bar f_3$     & 0.4     & 0.2    & $-0.6$ & $-1.4$ &  1.5     &   $-1.8$  \\
$f_t$         & 9.7  & 11.6 & 15.7 & 21.0 & 2.8  & 8.9  \\
$\bar f_t$  & $-2.2$     & $-2.4$    & $-3.1$    &  $-4.3$   &  18.3 &   18.9 \\
$f_l$         & $-11.8$    & $-12.6$   & $-15.7$   & $-18.8$   &  $-0.9$     &   $-11.5$    \\
$\bar f_l$  & 12.2 & 12.8 & 14.5 & 16.9 &  $-1.5$     &   14.1 \\
\hline
\end{tabular}
\label{forcepars}
\end{table}

For the force constants we used a simple parametrization, in which only central (bond-stretching) forces depend on the configuration, while the non-central force components are configuration-independent. We also limited the range of the force constants to second-nearest neighbors and assumed that the central force for bond $i-j$ depends only on the occupation of sites $i$ and $j$. The resulting model has 9 parameters including 6 central-force constants (2 per coordination sphere: for A-A, B-B, and A-B bonds), 2 non-central force constants for the nearest neighbors, and 1 isotropic non-central constant for the second-nearest neighbors. The fitted parameters are listed in Table \ref{fcs}. Note that the notation for
the first nearest neighbor force constants is given in a rotated reference frame to isolate the central forces. Specifically, for a (1/2,1/2,0) bond, the axes are rotated by $45^\circ$ around the $z$ axis so that the axis $x'$ lies along the (1,1,0) direction, parallel to the bond. $A_\mathrm{1RR}$ is the central force constant in this rotated frame. One can also write $A_\mathrm{1RR}=(A_\mathrm{1XX}+A_\mathrm{1YY})/2$, $A_\mathrm{1TT}=(A_\mathrm{1XX}-A_\mathrm{1YY})/2$.

\begin{table}[htb]
\caption{
Parameters of the cluster expansion for the force constants (units of dyn/cm). The notation for the first nearest neighbor constants is given in a reference frame
with direction $R$ parallel to the bond in $\hat x \hat y$ plane,  Z$\parallel \hat z$ as usual, and T orthogonal to both R and Z. 1RR and 2XX generate central forces.
}
\begin{tabular}{|c|c|r|r|r|r|r|r|}
\hline
\multicolumn{2}{|c|}{Pairs} & {AuFe} & {AuFe} & {AuFe} & {AuFe} & {AuFe$_2$} & {AuFe$_3$} \\
\multicolumn{2}{|c|}{$a$, \AA}          &   3.953 & 3.901  &   3.8  &   3.7  &  3.862   &  3.811    \\
\hline
\multirow{2}{*}{A-A}
 & 1RR & $8228$ & $11118$ & $17549$ & $25166$ & $13812$ & $17960$ \\
 & 2XX & $8859$ & $9596$ & $11356$ & $13667$ & $4766$ & $3459$ \\
\hline
\multirow{2}{*}{B-B}
 & 1RR & $58441$ & $68523$ & $91481$ & $119421$ & $73829$ & $84453$ \\
 & 2XX & $4449$ & $4455$ & $3928$ & $3548$ & $961$ & $-1768$ \\
\hline
\multirow{2}{*}{A-B}
 & 1RR & $25842$ & $31043$ & $42785$ & $57103$ & $35104$ & $41254$ \\
 & 2XX & $8272$ & $8869$ & $10019$ & $11146$ & $6144$ & $4874$ \\
\hline
\multirow{3}{*}{Any}
 & 1ZZ & $-4930$ & $-5862$ & $-7769$ & $-10065$ & $-3646$ & $-4935$ \\
 & 1TT & $-1922$ & $-2470$ & $-3691$ & $-5187$ & $-1566$ & $-1642$ \\
 & 2YY & $-1276$ & $-1247$ & $-1276$ & $-1399$ & $-1608$ & $-1664$ \\
\hline
\end{tabular}
\label{fcs}
\end{table}

The price paid for the accuracy of the CLDM is that the configuration-dependent force constant matrix can no longer be inverted in closed form, which hampers the direct calculation of the effective pair interactions in the random alloy. One approach, used previously in Ref.\ \onlinecite{CLDM}, is to fit the relaxation energy predicted by CLDM to an auxiliary many-body real-space cluster expansion, after which the definition (\ref{eq:Jderiv}) leads to
\begin{equation}
J^\mathrm{eff}_{ij} = \sum_{P\supset \{i,j\}}  \frac{m_P N^P_{ij}J_P\sigma_0^{N_P-2}}{m_{ij}},
\label{jeff}
\end{equation}
where $m_P$ is the multiplicity factor of cluster type $P$, $N_P$ the number of sites in $P$, $N^P_{ij}$ the number of edges of $P$ that are equivalent by symmetry to the pair $\{i,j\}$, $J_P$ the effective interaction for cluster type $P$, $m_{ij}$ the multiplicity factor of the pair cluster $\{i,j\}$, and $\sigma_0$ the concentration of the alloy.

Contrary to the CLDM, which captures the long-range singularity of the strain-induced interaction, the auxiliary CE has a finite range. Therefore, in Appendix \ref{s-cldm} we introduce another way to calculate $J_\mathrm{eff}(\mathbf{k})$ from CLDM, which allows one to retain its correct behavior at $k\to0$.

\section{Simplified CLDM (S-CLDM)}
\label{s-cldm}

The purpose of the ``simplified CLDM'' (S-CLDM) is to facilitate the calculation of the effective pair interactions $J_\mathrm{eff}({\bf k})$ in the random alloy while retaining the singular long-range part of the strain-induced interaction captured by the full CLDM. This long-range part may be particularly important when spinodal decomposition competes with ordering tendencies, as in the present case of the Au-Fe system.

We start by approximating the full CLDM by a S-CLDM with configuration-independent force constants. Within two shells of neighbors, as in the full CLDM (Table \ref{fcs}), the S-CLDM has 5 force constants (Table \ref{sfcs}). Given that the nearest-neighbor terms $f_1$ and $\bar f_1$ dominate in the Kanzaki force expansion (Table \ref{forcepars}), we only allow these two terms in the S-CLDM. The resulting S-CLDM has 7 parameters, but one of them is redundant, because the forces and force constants can be simultaneously scaled without changing the relaxation energies (\ref{erel}). We therefore fix $f_1$ to remove the degeneracy in the parameter space, which leaves 6 fitting parameters.

For each concentration and volume we calculate $E_{rel}$ from the corresponding CLDM for a set of 100 ordered structures, and then minimize the S-CLDM misfit using the coordinate-descent algorithm in the space of its 6 fitting parameters. To initialize this minimization process, we took the (configuration-independent) force constants fitted to the same first-principles data that were used to construct the full CLDM, while the starting values of $f_1$ and $\bar f_1$ were taken from the two-parameters fits in Table \ref{forcepars}. The final S-CLDM parameters are listed in Table \ref{sfcs}.

\begin{table}[htb]
\caption{Parameters of the cluster expansion in S-CLDM for the Kanzaki forces (meV/\AA) and force constants (dyn/cm).}
\begin{tabular}{|c|r|r|r|r|r|r|}
\hline
Pairs & {AuFe} & {AuFe} & {AuFe} & {AuFe} & {AuFe$_2$} & {AuFe$_3$} \\
$a$, \AA         &   3.953 & 3.901  &   3.8  &   3.7  &  3.862   &  3.811    \\
\hline
$f_1$& 252    &   295  &   398 &   536  &  310     &   333 \\
$\bar f_1$& 29    &   38  &   57 &   86  &  19     &   36 \\
\hline
1RR & $29758$ & $37050$ & $49993$ & $64932$ & $44347$ & $47507$ \\
2XX & $7202$ & $3828$ & $6388$ & $7795$ & $8520$ & $9840$ \\
1ZZ & $-6551$ & $-6514$ & $-8566$ & $-10937$ & $-19797$ & $-9581$ \\
1TT & $-980$ & $-582$ & $-1611$ & $-1560$ & $-4555$ & $-4703$\\
2YY & $-3539$ & $-4737$ & $-4341$ & $-1564$ & $-2121$ & $-6704$ \\
\hline
\end{tabular}
\label{sfcs}
\end{table}

As seen in Table \ref{fcs}, the force constants depend strongly on the configuration. Therefore, there is generally no reason to expect the S-CLDM with configuration-independent force constants to be adequate. However, Fig.\ \ref{fig:refit} demonstrates that the S-CLDM captures the dominant part of the full CLDM in the Au-Fe system.

\begin{figure}[htb]
\includegraphics[width=0.45\textwidth]{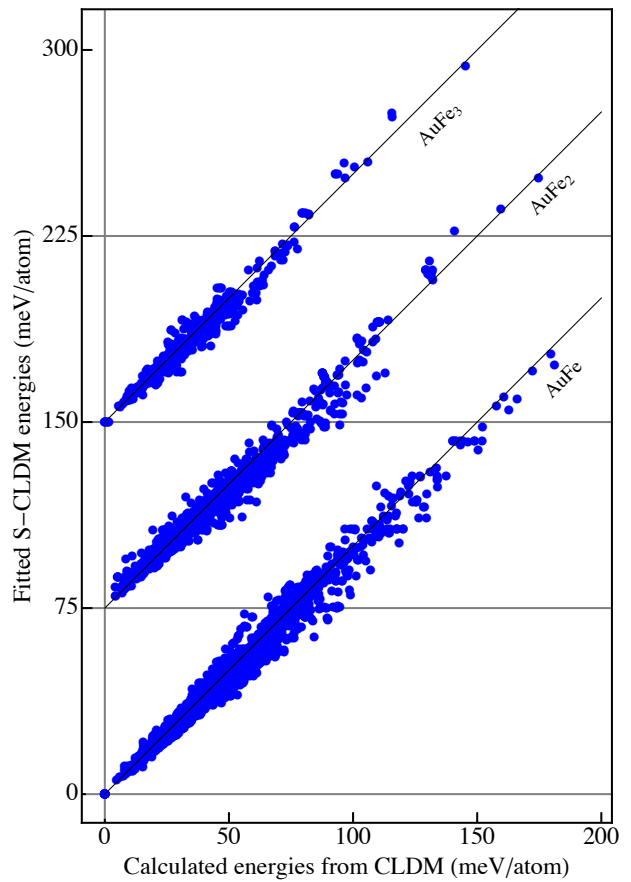}
\caption{
Relaxation energy (per atom) from S-CLDM \emph{vs} its input value from the full CLDM, for three concentrations. The set of structures is the same as the one used in the CE for $\Delta H_{rel}$, see Table \ref{tab:clexp}.
}
\label{fig:refit}
\end{figure}

Comparing the parameters in Tables \ref{sfcs} and \ref{forcepars}, we see that the $\bar f_1$ parameter in the optimal S-CLDM models is significantly reduced compared to the corresponding CLDM models. In the absence of $\bar f_1$ the interaction in S-CLDM would be pairwise, and the model would reduce to the simple Kanzaki-Krivoglaz-Khachaturyan model. The fact that a reasonably adequate S-CLDM has this property suggests that the effects of strong configuration dependence of the Kanzaki forces and force constants largely cancel each other in the Au-Fe system (and perhaps in other similar $3d$-$5d$ alloys). Thus, it may be reasonable to expect that the conventional Kanzaki-Krivoglaz-Khachaturyan model with parameters fitted directly to relaxation energies could perform well in such alloys, even though its \emph{effective} Kanzaki forces and force constants would have no relation to the physical forces and force constants.

The relatively small residual error of the S-CLDM (with respect to CLDM) is fitted to a real-space CE, the CV and misfits for which are listed in Table \ref{tab:clexp}. The contribution $J_\mathrm{res}(\mathbf{k})$ from this residual CE, as well as $J_{chem}(\mathbf{k})$, are calculated according to Eq.\ (\ref{jeff}). Finally, the dominant strain-induced contribution from S-CLDM is computed from Eq.\ (\ref{erel}). This is now straightforward, because the configuration-independent force constant matrix is easily inverted in reciprocal space. Taking second derivatives with respect to $\delta_i$ as in Eq.\ (\ref{eq:Jderiv}), we find, taking into account that $\sum_j\mathbf{e}_{ji}=0$ (see (\ref{fexp})):
\begin{equation}
J_\mathrm{eff}(\mathbf{k})=f^2\mathbf{e}(\mathbf{k})\hat A^{-1}(\mathbf{k})\mathbf{e}(\mathbf{k})+J_{chem}(\mathbf{k})+J_\mathrm{res}(\mathbf{k}),
\end{equation}
where $f=f_1+\sigma_0\bar f_1$, and $\mathbf{e}(\mathbf{k})$ is the Fourier transform of $\mathbf{e}_{ji}$.

\section{Proof that ordering striction has no effect on the ordering tendencies in a random alloy}
\label{app:proof}

The ordering tendencies are described by the effective interaction $J_\mathrm{eff}(\mathbf{k})$ (8) evaluated with respect to the disordered alloy. We consider a large disordered crystal or an average over an ensemble representing a disordered alloy. The quantities referring to the disordered configuration will be labeled by an index 0.

First, we prove a rather general statement. Consider a concentration wave, or any other particle-conserving inhomogeneity, in the disordered alloy, and let it be described by the parameters $\delta_i$ that were introduced above Eq.\ (\ref{eq:Jderiv}). (The disordered alloy can have short-range order; we only require that the deviation from it is fully determined by the local concentration changes $\delta_i$.) We further assume that the parent structure of the alloy is a Bravais lattice, such as the fcc lattice for Fe-Au alloys. Translational invariance then demands $\partial Q(\sigma)/\partial\delta_i=q$, where $Q(\sigma)$ is any function of configuration that is invariant under lattice translations, and $q$ is the same constant for all lattice sites $i$. To first order in $\delta_i$, we then have $\delta Q = q\sum_i\delta_i=0$ due to the conservation of the number of atoms. If the parent structure is not a Bravais lattice, the above conclusions apply to all inhomogeneities that preserve the numbers of atoms in each sublattice.

Now we turn to the ordering striction. In the main text, the ordering tendencies are studied both in the unstrained alloy and at a reduced lattice parameter, in order to examine the possible influence of a compressive strain in the core region of a nanoparticle. However, in reality it is not the volume that is fixed, but the pressure. Equilibrium of a system in an environment at (external) pressure $P_\mathrm{ext}$ corresponds to the minimum of the thermodynamic potential $\tilde H=E+P_\mathrm{ext}V$. In mechanical equilibrium $\tilde H$ is equal to the enthalpy $H$ of the system. The potential $\tilde H$ is a function of $P_\mathrm{ext}$, the occupation numbers $\sigma_i$, and the structural degrees of freedom $u_{\alpha\beta}$ and $\mathbf{w}_i$.

Let us define the strain tensor $u_{\alpha\beta}$ so that it vanishes in the equilibrium disordered alloy at pressure $P_0$ (whose volume is $V_0$). If a concentration wave in the disordered alloy induces a stress $\delta\sigma_{\alpha\beta}$ at $u_{\alpha\beta}=0$, the ordering striction contributes to the enthalpy (per unit volume) of a fully relaxed alloy as
\begin{equation}
H_{str} = -\frac12 \delta\sigma_{\alpha\beta}(\sigma)S_{\alpha\beta\gamma\delta}(\sigma)\delta\sigma_{\gamma\delta}(\sigma)
\label{Estrain}
\end{equation}
where $\hat S$ is the elastic compliance tensor.

The statement proved three paragraphs above applies to $\delta\sigma_{\alpha\beta}$. Therefore, $\delta\sigma_{\alpha\beta}$ is of second and $H_{str}$ of fourth order in $\delta_i$. Thus, $H_{str}$ has vanishing second derivatives in $\delta_i$ and does not contribute to $J_\mathrm{eff}(\mathbf{k})$. In other words, the ordering striction does not affect the ordering tendencies in the disordered alloy, and we are justified in using relaxation energy at constant \emph{strain} in Eq.\ (\ref{eq:DHchem}) instead of the relaxation enthalpy at given $P_\mathrm{ext}$.

On the other hand, $J_\mathrm{eff}(\mathbf{k})$ does depend on $P_\mathrm{ext}$. Indeed, we have:
\begin{align}
\frac{dH}{dP_\mathrm{ext}} = \left(\frac{\partial \tilde H}{\partial P_\mathrm{ext}}\right)_{u,w}=V_0(1+u_{\alpha\alpha}),
\end{align}
where the partial derivative is taken at fixed equilibrium values of $u_{\alpha\beta}$ and $\mathbf{w}_i$, and the derivatives are evaluated at $P_\mathrm{ext}=P_0$.
In the above, $u_{\alpha\alpha}=\delta V/V_0$ is the reduced volume relaxation of the given configuration $\sigma$ of the alloy with respect to the disordered alloy. This leads to
\begin{align}
\frac{dH}{dP_\mathrm{ext}}=1-S_{\alpha\alpha\gamma\delta}\delta\sigma_{\gamma\delta},
\end{align}
where the configuration-dependent term is quadratic in $\delta_i$. Thus, a pressure change adds a linear contribution to $J_{\mathrm{eff}}(\mathbf{k})$, which is seen in Fig.\ 8.

Note that the force constants depend strongly on the lattice parameter (see Table VII), which violates the harmonic approximation. The above expressions should, therefore, be understood in the spirit of the quasi-harmonic approximation, with the elastic compliance tensor and the striction stress corresponding to the relaxed nearly-random alloy at the given pressure.

\section{Energetics of substitutional defects in dilute Fe and Au}
\label{app:defects}

Given that all input structures except pure Fe and Au had 6 atoms per cell or fewer, the accuracy of the CE-SF predictions may deteriorate in dilute alloys. On the other hand, in the dilute limit, the slope of the formation enthalpy line for the random alloy represents the dissolution enthalpy of an isolated impurity atom, and it can be calculated separately using large supercells. Such calculations can only be performed for dynamically stable lattices, because otherwise an impurity atom would break the symmetry and collapse the dynamically unstable equilibrium. In our case, this excludes alloys based on fcc Fe and bcc Au, which are dynamically unstable at $T=0$ in GGA, as is typical for elemental metals in the ``wrong'' crystal structures.\cite{MehlBain2004} Here we examine the well-defined dissolution enthalpies of Fe in fcc Au and of Au in bcc Fe.

Fig.\ \ref{fig:defects} shows the formation enthalpies of supercells of different sizes $L\times L\times L$ containing a single impurity atom, including corrections calculated for $L>2$ from the continuous linear elasticity theory.\cite{Varvenne2013} For Au in bcc Fe, the formation enthalpy depends weakly on the size of the supercell, especially if the elastic correction is included. However, the formation enthalpy for Fe in fcc Au continues to change in the largest supercells that we have considered, while the elastic correction has a negligible effect on it. This dependence on the supercell size is much stronger than typical for defects in good metals where the electrostatic interaction is effectively screened.

\begin{figure}[hbt]
\includegraphics[width=0.45\textwidth]{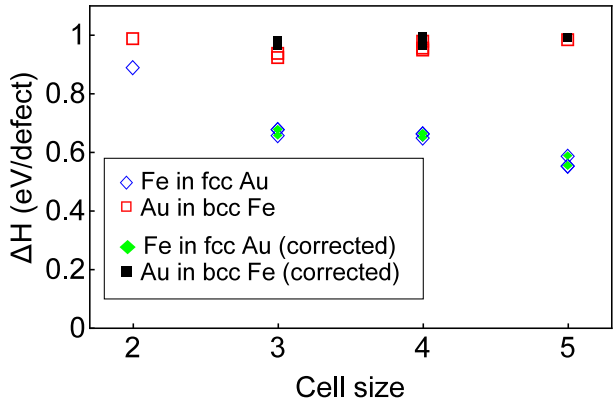}
\caption{Formation enthalpies of substitutional defects (diamonds: Fe in fcc Au matrix, squares: Au in bcc Fe matrix) vs. linear supercell size $L$.
Filled symbols: uncorrected values; open symbols: including the elastic corrections (only for $L>2$). Multiple symbols for the same $L$ correspond to different (but reasonably dense) $k$-point meshes.}
\label{fig:defects}
\end{figure}

Different data points in Fig.\ \ref{fig:defects} for the same $L$ were obtained using $k$-point meshes with densities in the range of  $20\times 20\times 20$ to $30\times 30\times 30$ (after mapping to the primitive cell), and we have always used equivalent $k$-point meshes for the pure element and the defect supercell. The discrepancy between these data points suggests that the defect formation enthalpy for Fe in fcc Au is not fully converged even for this rather dense $k$-point mesh, but this uncertainty is unlikely to be responsible for the observed size dependence.

We did not attempt to establish whether the supercell size dependence of the dissolution enthalpy of Fe in fcc Au reflects a simple concentration dependence or is sensitive to the specific geometric arrangement of the Fe atoms. Given the complexity of the Fermi surface of Au, which has narrow ``necks'' at the L points in the Brillouin zone, it is possible that a relatively small reduction of the Fermi momentum due to the small substitution of Fe is sufficient  to change the Fermi surface topology; this would lead to a largely concentration-dependent (rather than configuration-dependent) energetics. In this scenario, the data in Fig.\ \ref{fig:defects} suggests that the dissolution enthalpy of Fe in fcc Au rapidly increases with the concentration of Fe.

The dotted lines in Fig.\ \ref{fig:DeltaHunconstrained} of the main text indicate the slopes of the formation enthalpy in dilute random alloys corresponding to the dissolution enthalpies calculated here. For Au in bcc Fe, we use the converged value of 0.98 eV, and for Fe in fcc Au we took 0.57 eV corresponding to approximately 1\% Fe substitution. In both cases, these slopes agree reasonably well with the random-alloy lines predicted by the CEs.


\begin{thebibliography}{99}
\bibitem{note:symrelaxation} Only symmetry-conserving relaxations are considered here; some of the ``fully relaxed'' structures may thus correspond to saddle points in energy, provided the original symmetry forbids an important relaxation mode (such are, for example, pure bcc Au or pure HS fcc Fe at $T=0$).
\bibitem{note:DFTdetails} The PAW pseudopotentials treated only the unfilled top shells as valence electrons. The parametrization of Vosko, Wilk, and Nusair\cite{ref:VOSKOWN} was used with GGA-PBE. We used the basis set cut-off of 500 eV for the structural optimization and 350 eV for static calculations, and at least 16,000/$N_{at}$ k-points in the full Brilloin zone, where $N_{at}$ is the number of atoms in the unit cell.
\bibitem{note:relaxation}
Note that some structures shown in Fig.\ \ref{fig:ScoreVsX}, while stable within the original lattice, could still be dynamically unstable if the original geometry is highly symmetric [cf. Ref.\ \onlinecite{note:symrelaxation}].
\bibitem{note:Bain} We list the $c/a$ ratios using the fcc-derived definition of the $a$ lattice parameter. Technically, the structures along the Bain path have bct symmetry, and thus, should use $a_\text{bct}=a_\text{fcc}/\sqrt{2}$, so that the $c/a$ at the fcc positions becomes $c_\text{bct}/a_\text{bct}=\sqrt{2}$.
\bibitem{note:AFMBainSymmetry} The very shallow energy minimum and the apparently cubic (rather than linear) vanishing of the stress components near the G-type AFM minimum of \loz suggests a new, unusually high symmetry of the magnetic hamiltonian at this point; however, we found no physical justification for such an interpretation.
\bibitem{moduli} $E$ and $\nu$ were estimated by averaging the DFT elastic constants for \loz AuFe: $C_{11}=1.94$, $C_{22}=C_{33}=2.01$, $C_{12}=C_{13}=1.25$,
$C_{23}=1.32$, $C_{44}=0.76$, and $C_{55}=C_{66}=1.94$ Mbar.
\end{thebibliography}
\end{document}